\documentclass{andromedaone}

\usepackage{amsmath,amssymb}

\journal{BSM-2021}
\vol{Conference}
\jyear{Egypt}
\pages{Zewail City}

\received{xx March 2020}
\published{xx March 2020}

\def\be{\begin{equation}}
\def\ee{\end{equation}}
\def\bea{\begin{eqnarray}}
\def\eea{\end{eqnarray}}

\usepackage{slashed}
\usepackage{feynmp}
\usepackage{float}
\usepackage{graphicx}
\usepackage{array}
\usepackage{colortbl}
\usepackage{booktabs}
\usepackage{tcolorbox}
\newcolumntype{x}[1]{>{\centering\let\newline\\\arraybackslash\hspace{0pt}}p{#1}}

\usepackage{multicol}

\usepackage{fontawesome}
\usepackage{hyperref} 
\hypersetup{
     colorlinks = true,
     linkcolor = blue,
     anchorcolor = blue,
     citecolor = blue,
     filecolor = blue,
     urlcolor = blue,
     bookmarks=true,
     linktocpage=true
     }
     
\newcommand{\eps}{\varepsilon}
\newcommand{\vp}{H}
\newcommand{\tvp}{\widetilde{H}}

\newcommand{\vpj }{\mbox{${\vp^\dag i\,\raisebox{2mm}{\boldmath ${}^\leftrightarrow$}\hspace{-4mm} D_\mu\,\vp}$}}
\newcommand{\vpjt}{\mbox{${\vp^\dag i\,\raisebox{2mm}{\boldmath ${}^\leftrightarrow$}\hspace{-4mm} D_\mu^{\,I}\,\vp}$}}

\newcommand{\OO}{\ensuremath{\mathcal{O}}}

\renewcommand{\phi}{\ensuremath{\varphi}}

\newcommand{\sss}{\scriptscriptstyle}
\newcommand{\bpm}{\begin{pmatrix}}      
\newcommand{\epm}{\end{pmatrix}}

\newcommand{\Op}[1]{\OO_{\sss #1}}
\newcommand{\Opp}[2]{\OO_{\sss #1}^{\sss #2}}

\definecolor{lightgray}{rgb}{0.83, 0.83, 0.83}
\definecolor{lightpurp}{rgb}{0.901,0.796,0.882}

\newcommand{\lgc}{\cellcolor{lightgray} }

\def\wt{\widetilde}

\begin{document}

\title{ SMEFT Constraints on New Physics Beyond the Standard Model}

\author{John Ellis}
\address{Department of Physics, King's College London, London WC2R 2LS, United Kingdom;\\
National Institute of Chemical Physics \& Biophysics, Rävala 10, 10143 Tallinn, Estonia;\\
Theoretical Physics Department, CERN, CH-1211 Geneva 23, Switzerland}

\begin{abstract}
The Fermi effective theory of the weak interaction helped identify
the structure of the electroweak sector of the Standard Model, 
and the chiral effective Lagrangian pointed towards QCD as the theory of the strong interactions.
The Standard Model Effective Field Theory (SMEFT) is a systematic and model-independent framework
for characterizing experimental deviations from the predictions of the Standard Model and pointing
towards the structures of its possible extensions that is complementary to direct searches for
new physics beyond the Standard Model. This talk summarizes results from the first
global fit to data from LHC Run 2 and earlier experiments including dimension-6 SMEFT operators,
and gives examples how it can be used to constrain scenarios for new physics beyond the Standard
Model. In addition, some windows for probing dimension-8 SMEFT operators are also mentioned.
\end{abstract}

\maketitle

\begin{keyword}
Standard Model\sep Higgs\sep Effective Field Theory\sep Collider experiments\sep New physics\\
KCL-PH-TH/2021-035, CERN-TH-2021-083
\doi{10.31526/ACP}
\end{keyword}

\section{Introduction}

Since its formulation 54 years ago, many experiments have verified predictions of the Standard Model (SM).
The SU(3)$\times$SU(2)$\times$U(1) structure of its gauge interactions has been well established,
as have the representations and quantum numbers of the matter fermions. Predictions based on
these properties were tested below the 0.1\% level by experiments at LEP and other accelerators
before the LHC was put into operation~\cite{LEPEWWG}. However, there was no direct evidence for the
existence of the Higgs field, its effective potential and its Yukawa interactions with matter fermions
before LHC operations started. Following the discovery of the Higgs boson in 2012~\cite{Higgs}, these aspects of
the SM are now also being studied, and measured with increasing precision.

So far, all LHC measurements of cross sections for the production of SM particles are in agreement
with theoretical predictions, within the uncertainties in both theory and experiment. These include
the production of $W$ and $Z$ bosons alone or with multiple QCD jets, diboson pairs,
triple-boson combinations, $t \bar t$ pairs (also with a $\gamma, W$ or $Z$ boson), and single $t$ 
production alone or accompanied by a $\gamma, W$ or $Z$ boson~\cite{CMSplot}.

The SM has also been successful with its predictions for the production
of the Higgs boson by gluon collisions, vector-boson fusion, in association with a $W$ or $Z$, or
together with a $t \bar t$ pairs. The SM predictions for many Higgs decays have also been successful.
These have enabled the couplings of the Higgs boson to many SM particles, including the $\mu, \tau, b$,
$W$, $Z$ and $t$ to be measured. So far, the measurements are consistent 
with being proportional to the corresponding particle
masses, as expected in the SM~\cite{ATLASplot}. As we wrote a few years back "it walks and quacks like a Higgs boson"~\cite{Quacks}.

However, everything about the Higgs boson is puzzling. 
Consider the terms in the SM Lagrangian that involve the Higgs field:
\begin{equation}
    {\cal L} \; = \; y H \bar \psi \psi + \mu^2 |H|^2 \ \lambda |H|^4 - V_0 + \dots \, .
    \label{SMHL}
\end{equation}
The pattern of Yukawa couplings $y$ is completely mysterious, and intimately related to the flavour
problems of the SM: what is the origin of the hierarchies of fermion masses, the undeciphered matrix
of quark mixing and the origin of CP violation in the quark sector? The sign and magnitude of the
coefficient $\mu^2$ of the second term in (\ref{SMHL}) is equally baffling: in particular, how and why
is $|\mu|$ many orders of magnitude than the Planck scale? This is often termed the naturalness or
hierarchy problem. The magnitude of the quartic coupling $\lambda$ is also baffling. According to SM
calculations, $\lambda$ is so small that it would get renormalized to negative values at high scales~\cite{lambdaren} by the Higgs
coupling to the top quark, in which case the current electroweak vacuum would in principle be unstable,
and most of the Universe would not have reached it during the evolution of the Universe~\cite{notreached}. The $V_0$ 
term in (\ref{SMHL}) is equivalent to Einstein's cosmological constant or today's dark energy. This
receives contributions from spontaneous symmetry breaking and non-perturbative QCD effects that are
many orders of magnitude larger than the measured value. Here the mystery is not why $V_0 \ne 0$,
but rather why it is so small, over 120 orders of magnitude smaller than the ``natural"
gravitational scale. 

How should we search for new physics beyond the Standard Model?
This where the $\dots$ in (\ref{SMHL}) come in. There could be additional
interactions beyond those in the SM, of dimension 6 or more, which could be generated
by the exchanges of massive particles as yet undetected. These might involve the Higgs
field alone, or in combination with other SM fields, or only other fields. They should all be
analyzed together in order for the indirect search for new physics beyond the SM via higher-dimensional
interactions to be as powerful and model-independent as possible.

In looking for new physics, we may keep in mind the advice of the Chinese strategist Sun Tzu~\cite{ST}:\\
{\it ``...the direct method may be used...but indirect methods will be needed in order to
secure victory.... The direct and the indirect lead on to each other in turn. It is like moving in a circle....
Who can exhaust the possibilities of their combination?"\\}
This is the philosophy behind the use of the Standard Model Effective Field Theory
described in the following.

\section{The Standard Model Effective Field Theory (SMEFT)}

Effective Field Theories (EFTs) have a long and glorious history in particle physics.

Back in the 1930s, the interaction between radiation and electrons was described by an
embryonic form of QED, described by a dimension-4 Lagrangian term. However, the weak interactions
fell outside this framework, and Fermi proposed an effective four-fermion interaction to
describe them~\cite{Fermi}, described by some combination of operators of dimension 6. The natural
expectation was that this interaction was mediated by the exchange of massive
bosons. However, for many years the Lorentz structure of the four-fermion interaction was unclear: 
scalar/pseudoscalar, vector/axial vector or tensor, and hence also the spins of the exchanged bosons. 
Finally, it was established in the 1950s that the interaction was of $V - A$ form~\cite{FG}, so that the
intermediate boson(s) should have spin one. This observation pointed towards gauge theory, but
for a decade it was unknown how to give masses to the bosons. This problem was solved in the 1960s
by the discovery of the Englert-Brout-Higgs mechanism~\cite{EBH} and its application in the Standard Model.

In parallel during the 1930s, Yukawa proposed a theory of the strong nucleon-nucleon
force according to which it was mediated by mesons, of which the lightest was the pion that
was discovered in the 1940s. It was later discovered that the interactions of the pions involved
field derivatives, e.g., the dominant contribution to $\pi \pi$ scattering is an
interaction of the form $\partial \pi \partial \pi \pi \pi$. This structure is a manifestation
of an underlying chiral dynamics, which is an important clue to the nature of the structure of
the underlying forces between quarks. In particular, it implied that the exchanged gluons should
also be vector (rather than scalar) bosons, pointing the way towards QCD.

Will the EFT approach now provide us hints how to progress beyond the Standard Model? Many
extensions of the Standard Model postulate heavy particles whose exchanges could generate
effective interactions of dimensions higher than 4 whose structure may give valuable
information about the structure about extensions of the Standard Model. A first example is
provided by the light neutrino masses, which can be described by an effective interaction of
dimension 5~\cite{Weinberg} likely to be generated by the exchanges of massive fermions. The experimental fact that 
neutrino flavours mix in a very different way from quarks is presumably a consequence of
this underlying structure, and may give us a hint about new physics far above the electroweak
scale.

The Standard Model Effective Field Theory (SMEFT)~\cite{BW,Warsaw} provides a very powerful framework for analyzing in a global
way many LHC and other measurements. One assumes that the Standard Model assignments of the
quantum numbers of the Standard Model particles are correct, which determine the structure
of the dimension $\le 4$ terms in the Lagrangian, but accepts that it is incomplete. One
allows for additional higher-order interactions between the known particles that respect the
symmetries of the Standard Model, and
constrains their strengths via a joint analysis of Higgs data together with electroweak
precision measurements, diboson~\cite{EMSY} and top data~\cite{EMMSY}. This analysis can then be used to look in a 
general and model-independent way for possible physics beyond the Standard Model.

There are 2499 possible operators ${\cal O}_i$ in the dimension-6 part of the SMEFT Lagrangian:
\begin{equation}
    {\cal L}_{\rm SMEFT} \; = {\cal L}_{\rm SM} + \sum_{i = 1}^{2499} \frac{C_i}{\Lambda^2} {\cal O}_i \, ,
\end{equation}
where $\Lambda$ is the new physics scale, which is assumed to be much larger than the masses of
the Standard Model particles and the kinematic scales of the processes studied, and the $C_i$
are operator coefficients whose magnitudes depend, e.g., whether the underlying massive
particles are weakly- or strongly-interacting. It is clearly impractical to include 2499 parameters
in a global fit, so one generally makes simplifying assumptions about the flavour
structure of the operators involving fermions. In the following we consider two possibilities~\cite{EMMSY}:
a flavour-universal scenario in which their coefficients have an SU(3)$^5$ symmetry, and a
top-specific scenario in which the symmetry is reduced to SU(2)$^2 \times$SU(3)$^3$. When making our global fit,
we treat $G_F, M_Z$ and the fine structure constant $\alpha$ as input parameters, and work to
linear order in the operator coefficients, i.e, we consider only the interferences of the 
${\cal O}_i$ with the Standard Model interactions. Thus, we work to order
$1/\Lambda^2$ and neglect consistently dimension-8 interactions that would contribute at ${\cal O}(1/\Lambda^4)$. 

The Table below lists all the relevant dimension-6 SMEFT operators, including those containing
two and four fermions, which are important when considering top observables. The grey cells 
contain operators that violate SU(3)$^5$ universality. If we impose universality, the
following 20 operators affect primarily electroweak precision observables,
bosonic observables and Yukawa couplings, respectively:
\begin{align}
    \text{EWPO: } &\quad \mathcal{O}_{HWB} \, , \, \mathcal{O}_{HD} \, , \, \mathcal{O}_{ll} \, , \, \mathcal{O}_{Hl}^{(3)} \, , \, \mathcal{O}_{Hl}^{(1)} \, , \, \mathcal{O}_{He} \, , \, \mathcal{O}_{Hq}^{(3)} \, , \, \mathcal{O}_{Hq}^{(1)} \, , \, \mathcal{O}_{Hd} \, , \, \mathcal{O}_{Hu} \, , \, \nonumber \\
    \text{Bosonic: } &\quad  \mathcal{O}_{H\Box} \, , \, \mathcal{O}_{HG} \, , \, \mathcal{O}_{HW} \, , \, \mathcal{O}_{HB} \, , \, \mathcal{O}_{W} \, , \, \mathcal{O}_G \, , \, \nonumber \\
    \text{Yukawa: } &\quad \mathcal{O}_{\tau H} \, , \, \mathcal{O}_{\mu H} \, , \, \mathcal{O}_{bH} \, , \, \mathcal{O}_{tH} \, .
\label{eq:flavunivops}
\end{align}
Another 14 two- and four-fermion operators appear when we break the flavour universality down to 
SU(2)$^2 \times$SU(3)$^3$ in the top-specific scenario:
\begin{align}
    \label{eq:topspecificops}
    \text{Top 2F: } &\quad \mathcal{O}_{HQ}^{(3)} \, , \, \mathcal{O}_{HQ}^{(1)} \, , \, \mathcal{O}_{Ht} \, , \, \mathcal{O}_{tG} \, , \, \mathcal{O}_{tW} \, , \, \mathcal{O}_{tB} \, , \nonumber \\
    \text{Top 4F: } &\quad \mathcal{O}_{Qq}^{3,1} \, , \, \mathcal{O}_{Qq}^{3,8} \, , \, \mathcal{O}_{Qq}^{1,8} \, , \, \mathcal{O}_{Qu}^{8} \, , \, \mathcal{O}_{Qd}^{8} \, , \, \mathcal{O}_{tQ}^{8} \, , \, \mathcal{O}_{tu}^{8} \, , \, \mathcal{O}_{td}^{8} \, .
\end{align}
Results from global fits under these two flavour assumptions are discussed in the following Section.

\begin{table}[t] 
\centering
\hspace{1mm}
\renewcommand{\arraystretch}{1.0}
\scalebox{1.1}{
\begin{tabular}{||c|c||c|c||c|c||} 
\hline \hline
\multicolumn{2}{||c||}{$X^3$} & 
\multicolumn{2}{|c||}{$\vp^6$~ and~ $\vp^4 D^2$} &
\multicolumn{2}{|c||}{$\psi^2\vp^3$}\\
\hline
$\Op{G}$                 & $f^{ABC} G_\mu^{A\nu} G_\nu^{B\rho} G_\rho^{C\mu} $ &  
$\Op{\vp}$               & $(\vp^\dag\vp)^3$ &
\lgc $\Op{e\vp}$           & \lgc $(\vp^\dag \vp)(\bar l_p e_r \vp)$\\
$\Op{\wt G}$          & $f^{ABC} \wt G_\mu^{A\nu} G_\nu^{B\rho} G_\rho^{C\mu} $ &   
$\Op{\vp\Box}$        & $(\vp^\dag \vp)\raisebox{-.5mm}{$\Box$}(\vp^\dag \vp)$ &
\lgc $\Op{u\vp}$           & \lgc $(\vp^\dag \vp)(\bar q_p u_r \tvp)$\\
$\Op{W}$                 & $\eps^{IJK} W_\mu^{I\nu} W_\nu^{J\rho} W_\rho^{K\mu}$ &    
$\Op{\vp D}$          & $\left(\vp^\dag D^\mu\vp\right)^\star \left(\vp^\dag D_\mu\vp\right)$ &
\lgc $\Op{d\vp}$           & \lgc $(\vp^\dag \vp)(\bar q_p d_r \vp)$\\
$\Op{\wt W}$          & $\eps^{IJK} \wt W_\mu^{I\nu} W_\nu^{J\rho} W_\rho^{K\mu}$ &&&& \\    
\hline \hline
\multicolumn{2}{||c||}{$X^2\vp^2$} &
\multicolumn{2}{|c||}{$\psi^2 X\vp$} &
\multicolumn{2}{|c||}{$\psi^2\vp^2 D$}\\ 
\hline
$\Op{\vp G}$            & $\vp^\dag \vp\, G^A_{\mu\nu} G^{A\mu\nu}$ & 
\lgc $\Op{eW}$               & \lgc $(\bar l_p \sigma^{\mu\nu} e_r) \tau^I \vp W_{\mu\nu}^I$ &
$\Opp{\vp l}{(1)}$      & $(\vpj)(\bar l_p \gamma^\mu l_r)$\\
$\Op{\vp\wt G}$         & $\vp^\dag \vp\, \wt G^A_{\mu\nu} G^{A\mu\nu}$ &  
\lgc $\Op{eB}$               & \lgc $(\bar l_p \sigma^{\mu\nu} e_r) \vp B_{\mu\nu}$ &
$\Opp{\vp l}{(3)}$      & $(\vpjt)(\bar l_p \tau^I \gamma^\mu l_r)$\\
$\Op{\vp W}$            & $\vp^\dag \vp\, W^I_{\mu\nu} W^{I\mu\nu}$ & 
\lgc $\Op{uG}$               & \lgc $(\bar q_p \sigma^{\mu\nu} T^A u_r) \tvp\, G_{\mu\nu}^A$ &
$\Op{\vp e}$            & $(\vpj)(\bar e_p \gamma^\mu e_r)$\\
$\Op{\vp\wt W}$         & $\vp^\dag \vp\, \wt W^I_{\mu\nu} W^{I\mu\nu}$ &
\lgc $\Op{uW}$               & \lgc $(\bar q_p \sigma^{\mu\nu} u_r) \tau^I \tvp\, W_{\mu\nu}^I$ &
$\Opp{\vp q}{(1)}$      & $(\vpj)(\bar q_p \gamma^\mu q_r)$\\
$\Op{\vp B}$            & $ \vp^\dag \vp\, B_{\mu\nu} B^{\mu\nu}$ &
\lgc $\Op{uB}$               &\lgc  $(\bar q_p \sigma^{\mu\nu} u_r) \tvp\, B_{\mu\nu}$&
$\Opp{\vp q}{(3)}$      & $(\vpjt)(\bar q_p \tau^I \gamma^\mu q_r)$\\
$\Op{\vp\wt B}$         & $\vp^\dag \vp\, \wt B_{\mu\nu} B^{\mu\nu}$ &
\lgc $\Op{dG}$               & \lgc $(\bar q_p \sigma^{\mu\nu} T^A d_r) \vp\, G_{\mu\nu}^A$ & 
$\Op{\vp u}$            & $(\vpj)(\bar u_p \gamma^\mu u_r)$\\
$\Op{\vp WB}$           & $ \vp^\dag \tau^I \vp\, W^I_{\mu\nu} B^{\mu\nu}$ &
\lgc $\Op{dW}$               & \lgc $(\bar q_p \sigma^{\mu\nu} d_r) \tau^I \vp\, W_{\mu\nu}^I$ &
$\Op{\vp d}$            & $(\vpj)(\bar d_p \gamma^\mu d_r)$\\
$\Op{\vp\wt WB}$        & $\vp^\dag \tau^I \vp\, \wt W^I_{\mu\nu} B^{\mu\nu}$ &
\lgc $\Op{dB}$               &\lgc  $(\bar q_p \sigma^{\mu\nu} d_r) \vp\, B_{\mu\nu}$ &
\lgc $\Op{\vp u d}$          & \lgc $i(\tvp^\dag D_\mu \vp)(\bar u_p \gamma^\mu d_r)$\\
\hline \hline
\hline\hline
\multicolumn{2}{||c||}{$(\bar LL)(\bar LL)$} & 
\multicolumn{2}{|c||}{$(\bar RR)(\bar RR)$} &
\multicolumn{2}{|c||}{$(\bar LL)(\bar RR)$}\\
\hline
$\Op{ll}$               & $(\bar l_p \gamma_\mu l_r)(\bar l_s \gamma^\mu l_t)$ &
$\Op{ee}$               & $(\bar e_p \gamma_\mu e_r)(\bar e_s \gamma^\mu e_t)$ &
$\Op{le}$               & $(\bar l_p \gamma_\mu l_r)(\bar e_s \gamma^\mu e_t)$ \\
$\Opp{qq}{(1)}$  & $(\bar q_p \gamma_\mu q_r)(\bar q_s \gamma^\mu q_t)$ &
$\Op{uu}$        & $(\bar u_p \gamma_\mu u_r)(\bar u_s \gamma^\mu u_t)$ &
$\Op{lu}$               & $(\bar l_p \gamma_\mu l_r)(\bar u_s \gamma^\mu u_t)$ \\
$\Opp{qq}{(3)}$  & $(\bar q_p \gamma_\mu \tau^I q_r)(\bar q_s \gamma^\mu \tau^I q_t)$ &
$\Op{dd}$        & $(\bar d_p \gamma_\mu d_r)(\bar d_s \gamma^\mu d_t)$ &
$\Op{ld}$               & $(\bar l_p \gamma_\mu l_r)(\bar d_s \gamma^\mu d_t)$ \\
$\Opp{lq}{(1)}$                & $(\bar l_p \gamma_\mu l_r)(\bar q_s \gamma^\mu q_t)$ &
$\Op{eu}$                      & $(\bar e_p \gamma_\mu e_r)(\bar u_s \gamma^\mu u_t)$ &
$\Op{qe}$               & $(\bar q_p \gamma_\mu q_r)(\bar e_s \gamma^\mu e_t)$ \\
$\Opp{lq}{(3)}$                & $(\bar l_p \gamma_\mu \tau^I l_r)(\bar q_s \gamma^\mu \tau^I q_t)$ &
$\Op{ed}$                      & $(\bar e_p \gamma_\mu e_r)(\bar d_s\gamma^\mu d_t)$ &
$\Opp{qu}{(1)}$         & $(\bar q_p \gamma_\mu q_r)(\bar u_s \gamma^\mu u_t)$ \\ 
&& 
$\Opp{ud}{(1)}$                & $(\bar u_p \gamma_\mu u_r)(\bar d_s \gamma^\mu d_t)$ &
$\Opp{qu}{(8)}$         & $(\bar q_p \gamma_\mu T^A q_r)(\bar u_s \gamma^\mu T^A u_t)$ \\ 
&& 
$\Opp{ud}{(8)}$                & $(\bar u_p \gamma_\mu T^A u_r)(\bar d_s \gamma^\mu T^A d_t)$ &
$\Opp{qd}{(1)}$ & $(\bar q_p \gamma_\mu q_r)(\bar d_s \gamma^\mu d_t)$ \\
&&&&
$\Opp{qd}{(8)}$ & $(\bar q_p \gamma_\mu T^A q_r)(\bar d_s \gamma^\mu T^A d_t)$\\
\hline\hline
\multicolumn{2}{||c||}{ $(\bar LR)(\bar RL)$ and $(\bar LR)(\bar LR)$} &
\multicolumn{4}{|c||}{ $B$-violating}\\\hline
\lgc $\Op{ledq}$  & \lgc $(\bar l_p^j e_r)(\bar d_s q_t^j)$ &
\lgc $\Op{duq}$   &  \multicolumn{3}{|c||}{\lgc $\eps^{\alpha\beta\gamma} \eps_{jk} 
 \left[ (d^\alpha_p)^T C u^\beta_r \right]\left[(q^{\gamma j}_s)^T C l^k_t\right]$}\\
\lgc $\Opp{quqd}{(1)}$ & \lgc $(\bar q_p^j u_r) \eps_{jk} (\bar q_s^k d_t)$ &
\lgc $\Op{qqu}$ & \multicolumn{3}{|c||}{\lgc $\eps^{\alpha\beta\gamma} \eps_{jk} 
  \left[ (q^{\alpha j}_p)^T C q^{\beta k}_r \right]\left[(u^\gamma_s)^T C e_t\right]$}\\
\lgc $\Opp{quqd}{(8)}$ & \lgc $(\bar q_p^j T^A u_r) \eps_{jk} (\bar q_s^k T^A d_t)$ &
\lgc $\Op{qqq}$ & \multicolumn{3}{|c||}{ \lgc $\eps^{\alpha\beta\gamma} \eps_{jn} \eps_{km} 
  \left[ (q^{\alpha j}_p)^T C q^{\beta k}_r \right]\left[(q^{\gamma m}_s)^T C l^n_t\right]$}\\
\lgc $\Opp{lequ}{(1)}$ &\lgc  $(\bar l_p^j e_r) \eps_{jk} (\bar q_s^k u_t)$ &
\lgc $\Op{duu}$ & \multicolumn{3}{|c||}{ \lgc $\eps^{\alpha\beta\gamma} 
  \left[ (d^\alpha_p)^T C u^\beta_r \right]\left[(u^\gamma_s)^T C e_t\right]$}\\
\lgc $\Opp{lequ}{(3)}$ & \lgc $(\bar l_p^j \sigma_{\mu\nu} e_r) \eps_{jk} (\bar q_s^k \sigma^{\mu\nu} u_t)$ &
& \multicolumn{3}{|c||}{}\\
\hline\hline
\end{tabular}
} 
{\it Dimension-6 operators considered in~\cite{EMMSY}. 
The operators in cells shaded grey break flavour SU(3)$^5$ explicitly. \label{tab:operators}} 
\end{table}

\section{Global SMEFT Fits to to Electroweak, Diboson, Higgs and Top Data}

\begin{figure}[t] 
\centering
\includegraphics[width=0.6\textwidth]{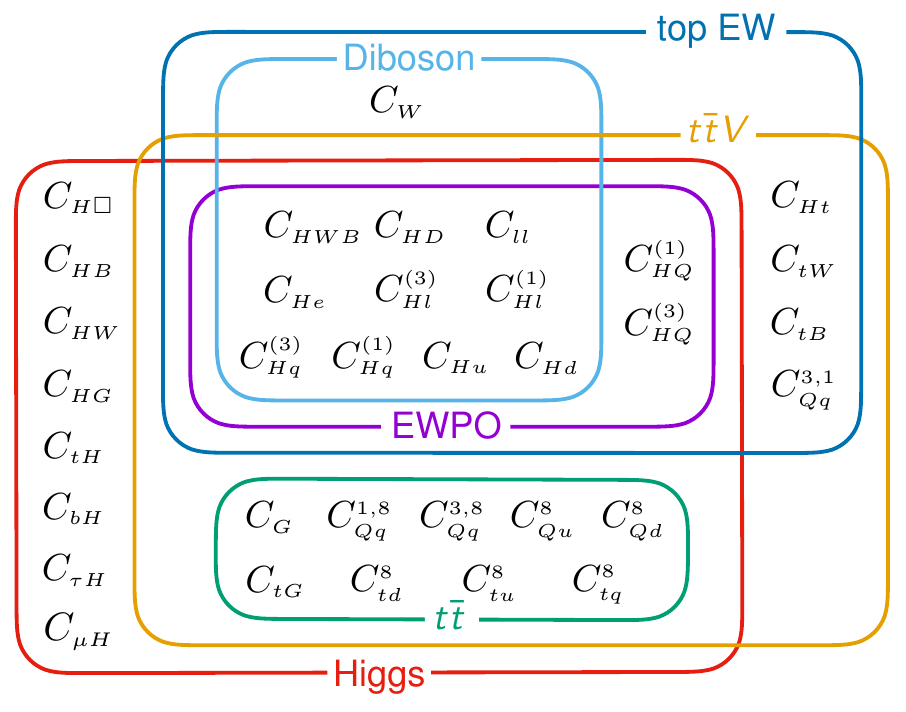}
\caption{
\label{fig:venn}
\it An illustration~\cite{EMMSY} how various classes of observables depend on the 34 dimension-6 operators (\ref{eq:flavunivops}, \ref{eq:topspecificops}).} 
\end{figure}

This Section presents some of the results from global fits to the dimension-6 operators listed above,
using precision electroweak data and $W^+ W^-$ data from LEP, as well as Higgs, diboson and 
top data from Runs 1 and 2 of the LHC: please see~\cite{EMMSY} for the complete results.
Fig.~\ref{fig:venn} shows schematically how the different operators contribute to various
sets of measurements. It emphasizes their interdependence, and hence the need for a
global analysis taking all the data sets into consideration. Overall, we included 328 individual
data points in our analysis, including separate bins characterizing kinematic variables in many
differential distributions.

\begin{figure}[h!] 
\centering
\includegraphics[width=0.6\textwidth]{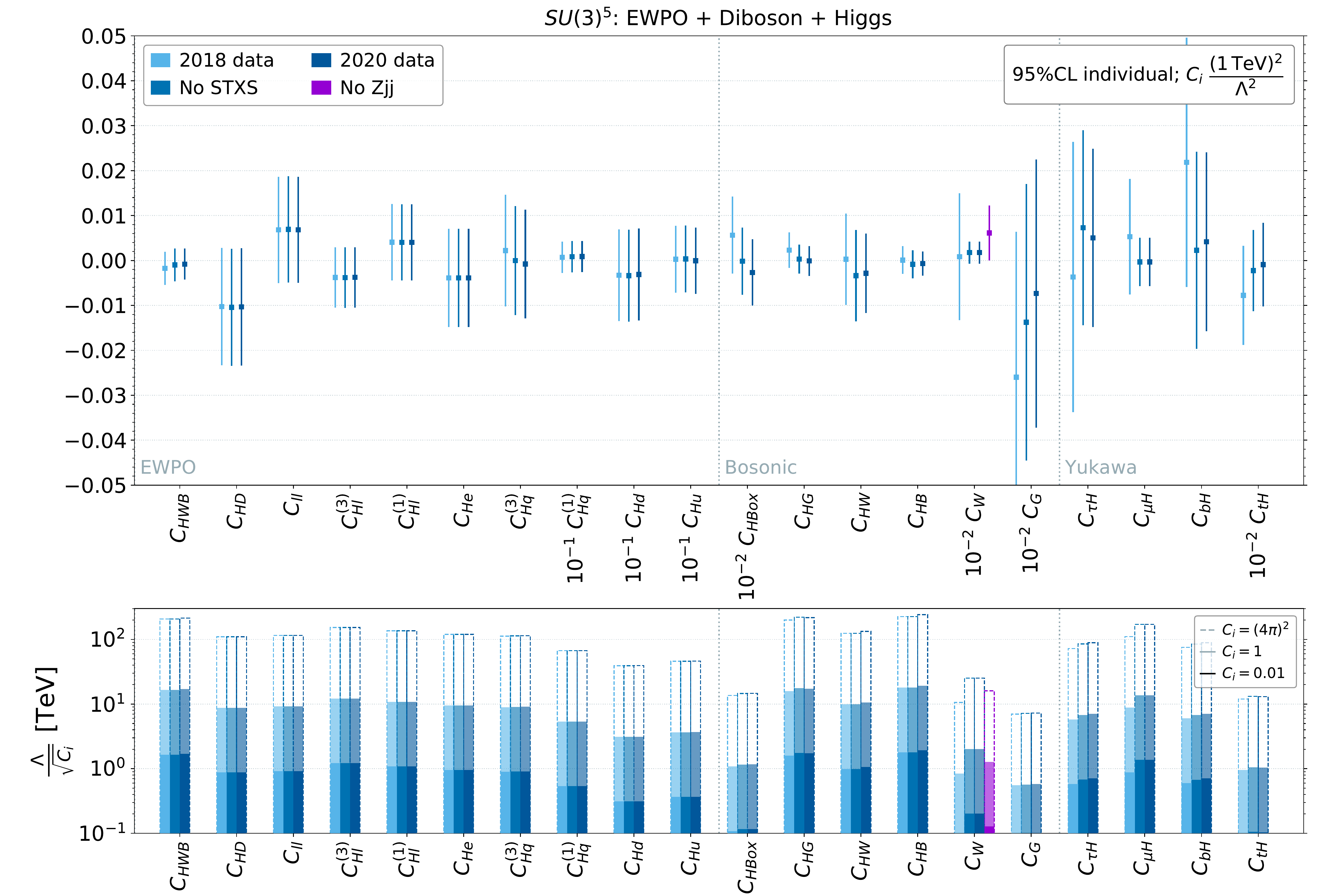}\\
\includegraphics[width=0.6\textwidth]{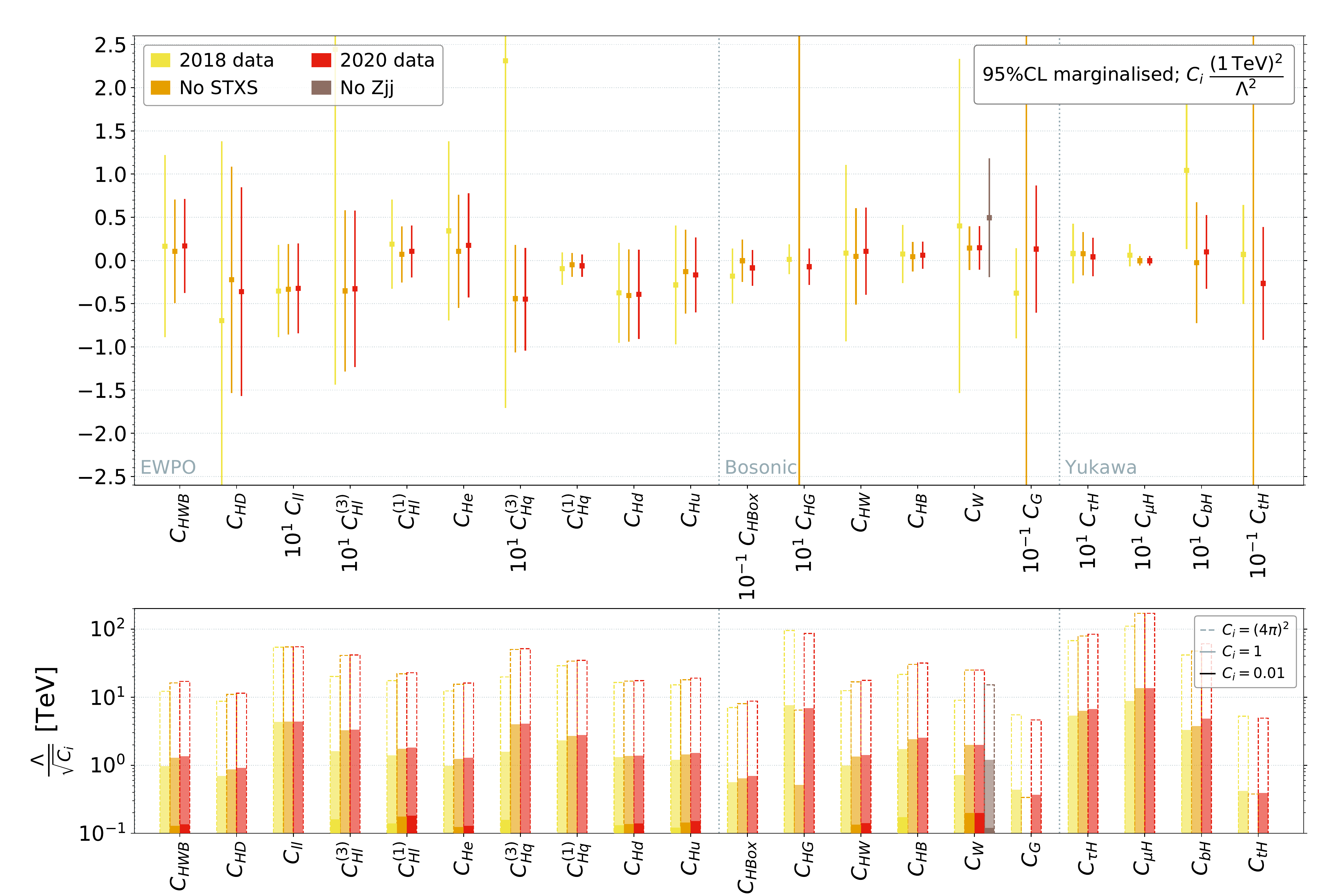}\\
\caption{\it Constraints at the 95\% confidence level in the
flavour-universal SU(3)$^5$ scenario on individual and marginalized 
operator coefficients 
$C_i (1~{\rm TeV})^2/\Lambda^2$ (top and third panels) and the corresponding scales $\Lambda$ for the indicated values of the ${C_i}$
(second and bottom panels), from a combined linear fit to the Higgs, diboson and electroweak precision observables using the
data sets indicated by colour coding~\cite{EMMSY}.}
\label{fig:flavour_universal_EWdiBH}
\end{figure}

Fig.~\ref{fig:flavour_universal_EWdiBH} shows the overall results from our flavour-universal
SU(3)$^5$-symmetric fit. The upper pair of panels shows the constraints we obtain on
the operators when they are each switched on individually, whereas in the lower pair
of panels all the 20 operators are switched on simultaneously, and the constraints on the
operators are those obtained by marginalizing over the coefficients of all the other operators.
The upper panel in each pairs shows the constraints on the numerical coefficients $C_i$ scaled for
an operator scale $\Lambda = 1$~TeV, and the lower panel in each pair shows the constraints
on $\Lambda$ obtained for various indicated values of the $C_i$. The colours in the histograms
show the effects of including only Run-1 Higgs data, also Run-2 data, kinematic information
denoted by STXS~\cite{STXS}, and dropping data on $Z$ + dijet data.

In the case of the individual fit
we see that the obtained constraints on $\Lambda$ are all in the multi-TeV range in the
strong-coupling limit $C_i = (4 \pi)^2$, in which case whereas the linear treatment of the EFT
approximation used here should be quite accurate. For $C_i = 1$, the individual limits are still
$\gtrsim 1$~TeV, except for $C_G$, which is better constrained by a quadratic treatment of
multi-jet data. On the other hand, several of the constraints on other operator scales get
weaker than 1~TeV in the weak-coupling case $C_i = 0.01$. Also we see in the lower panels of
Fig.~\ref{fig:flavour_universal_EWdiBH} that all the constraints are significantly weaker
in the marginalized fit, as could be expected in a fit with more parameters.

\begin{figure}[t] 
\centering
\includegraphics[width=0.6\textwidth]{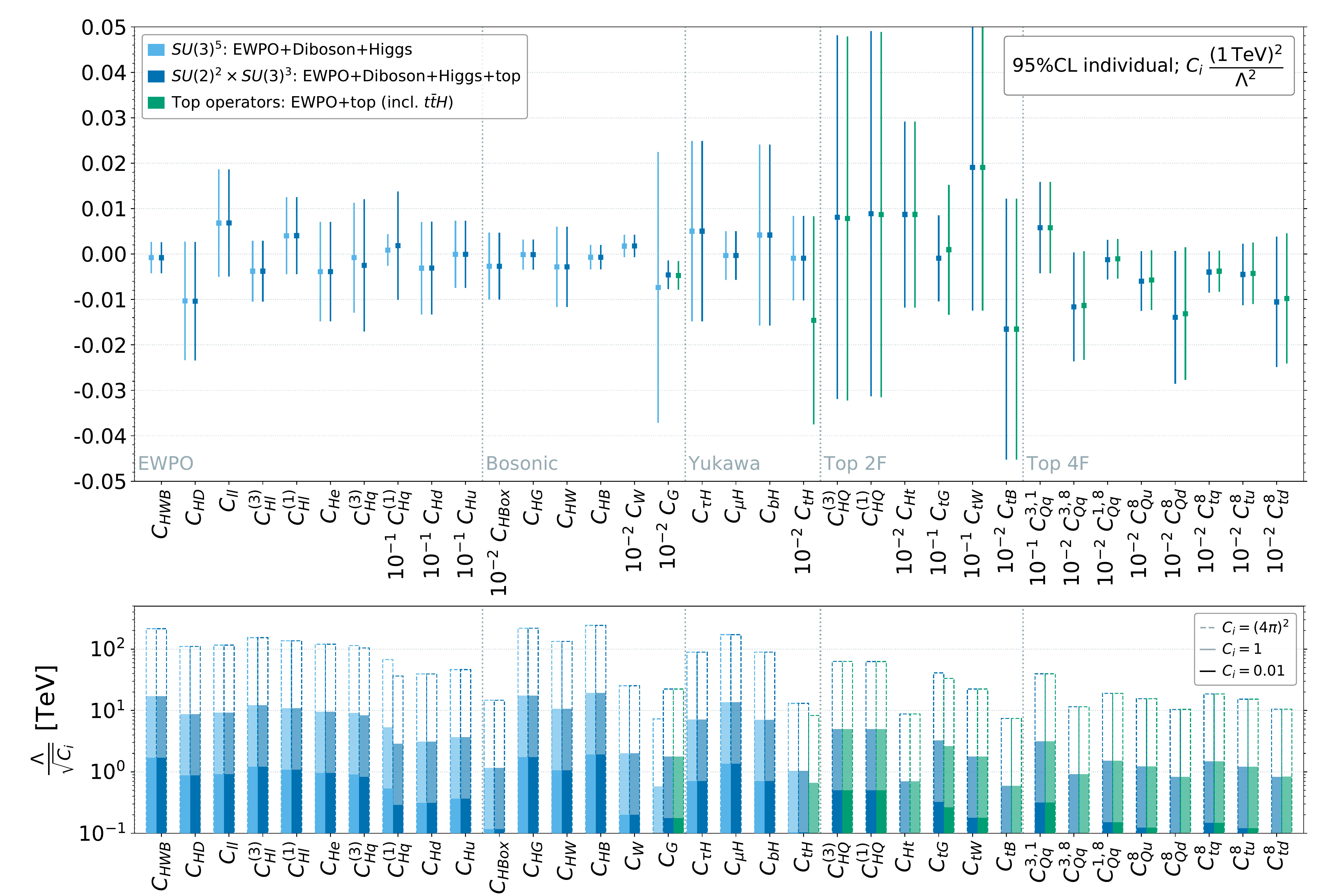}\\
\includegraphics[width=0.6\textwidth]{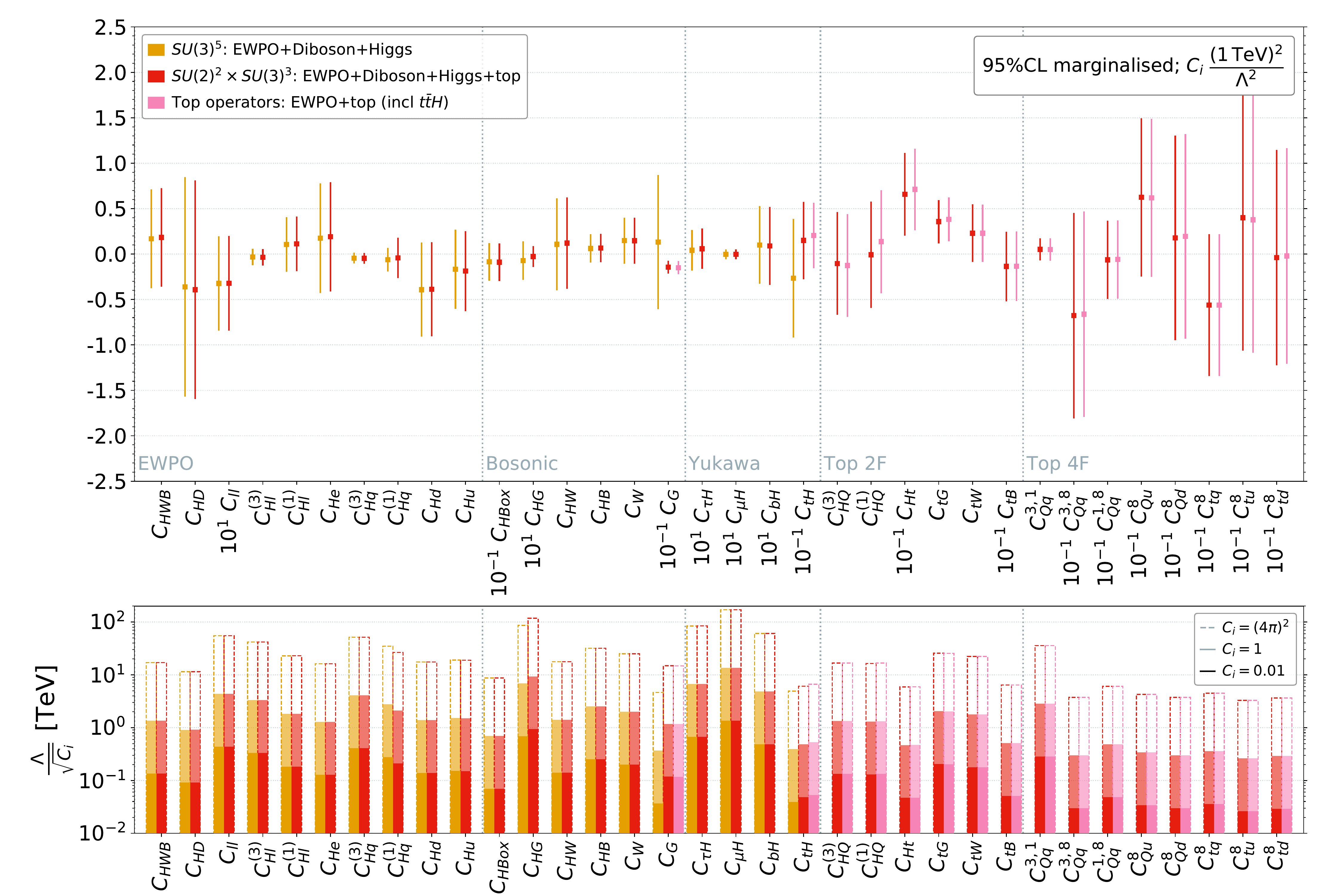}
\caption{\it \label{fig:all_fit} Constraints at the 95\% confidence level in the
top-specific SU(2)$^2\times$SU(3)$^3$ scenario on 
individual and marginalized operator coefficients 
$C_i (1~{\rm TeV})^2/\Lambda^2$ (top and third panels) and the corresponding scales $\Lambda$ for the indicated values of the ${C_i}$
(second and bottom panels), from a combined linear fit to the Higgs, diboson and electroweak precision observables using the
data sets indicated by colour coding~\cite{EMMSY}.}
\end{figure}

Fig.~\ref{fig:all_fit} shows corresponding results for the top-specific 
SU(2)$^2 \times$SU(3)$^3$-symmetric fit with 34 operators. We see that the constraints
in the electroweak, bosonic and Yukawa sectors are similar to those in the universal
SU(3)$^5$-symmetric fit. However, the constraints in the two- and four-fermion
sectors are generally weaker. The constraints on $\Lambda$ are generally $\gtrsim 10$~TeV
in the individual strong-coupling fit, but fall to $\gtrsim 300$~GeV in the marginalized
fit with $C_i = 1$. The linear EFT approximation is likely to break down in the top sector
for the weak-coupling case $C_i = 0.01$.~\footnote{For results in the quadratic EFT
approximation assuming that the precision electroweak data agree perfectly with the Standard Model
and ignoring possible dimension-8 operator contributions, see~\cite{SMEFit}.}

\begin{figure}[htb]! 
\centering
\includegraphics[width=0.6\textwidth]{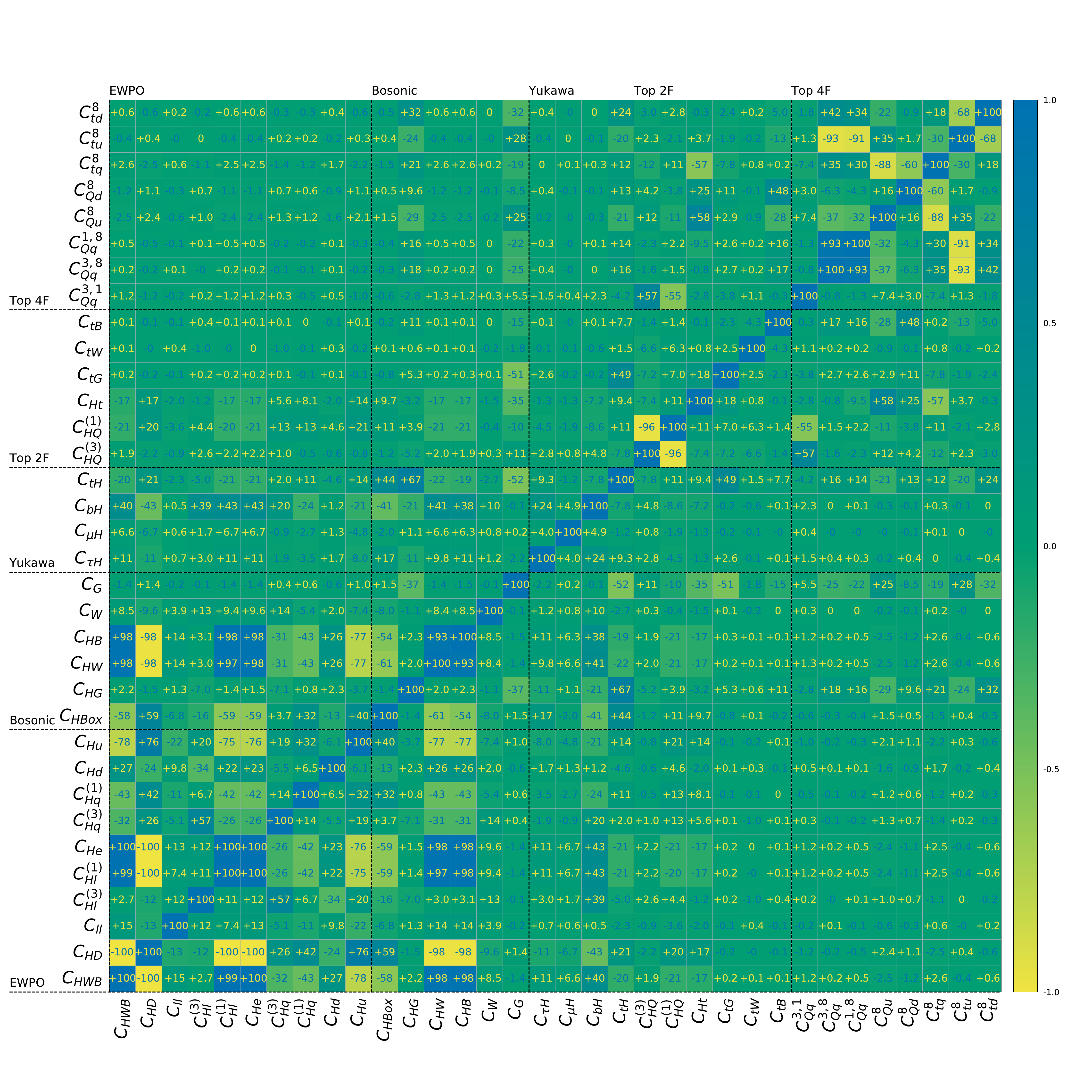}
\caption{\it The correlation matrix for the 34 operator coefficients in
the marginalized top-specific fit~\cite{EMMSY}. The entries are colour-coded 
according to the magnitudes of the correlation coefficients.
}
\label{fig:correlationmatrixfull}
\end{figure}

As seen in Fig.~\ref{fig:venn}, many SMEFT operators
contribute to the same data sectors, introducing correlations between the constraints on
their coefficients. Moreover, many also contribute to multiple data sectors,
introducing correlations between sectors. Fig.~\ref{fig:correlationmatrixfull} displays
the full 34$\times$34 correlation matrix for the operator coefficients in the marginalized
top-specific fit. We see substantial operator correlations within the EWPO and bosonic sectors,
and also within the two- and four-fermion top sectors, as could be expected.
However, correlations between operators in these sectors and in the Yukawa sector can also
be significant, as reflected in the numerical values in off-diagonal squares in the matrix.

\begin{figure}[htb]!
\hspace{-10mm}
\centering
\hspace{-10mm}
\includegraphics[width=0.6\textwidth]{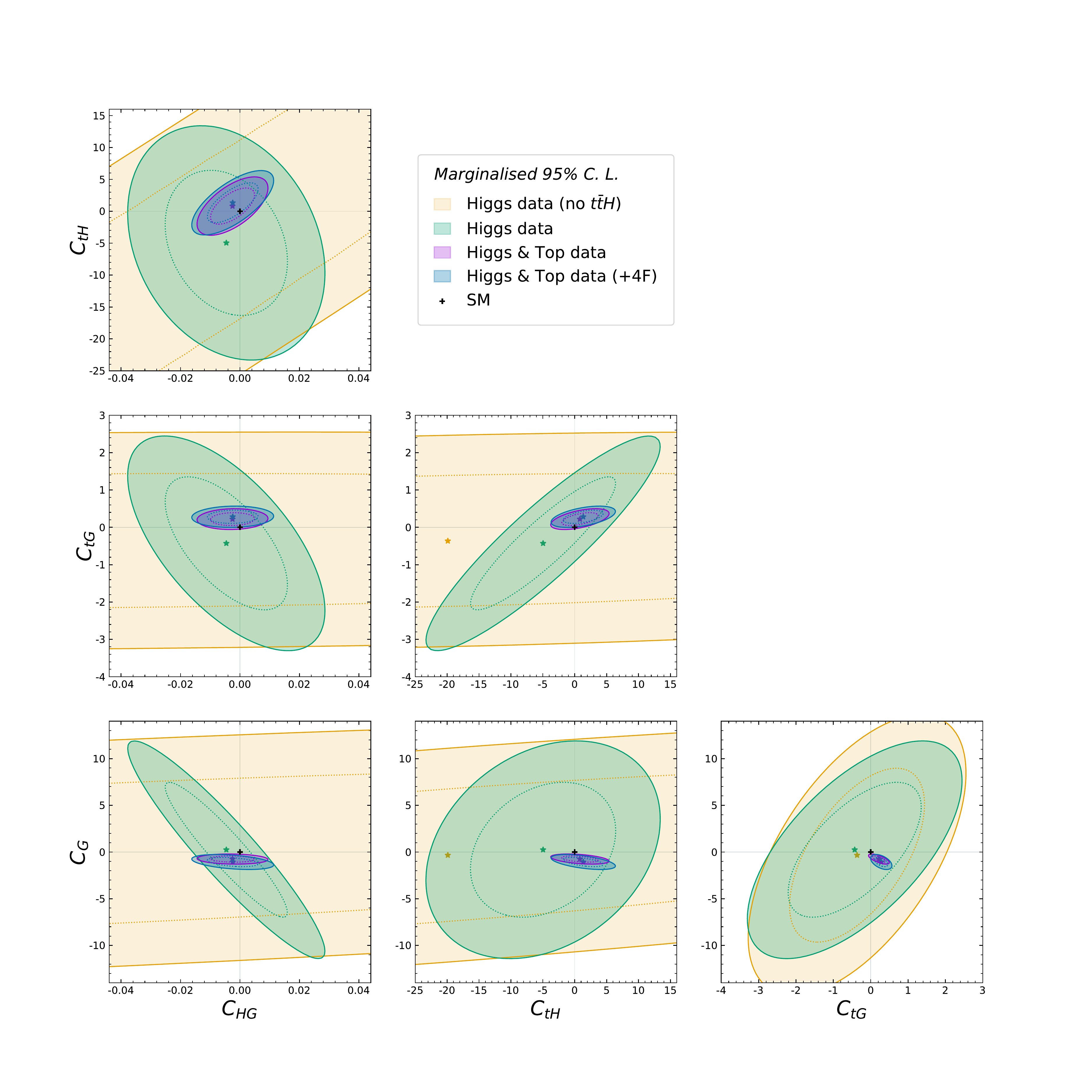}
\caption{\it \label{fig:interplay1} Constraints on pairs of operator coefficients 
at the 95\% confidence level, marginalized over the other 
degrees of freedom in the `Higgs only' operator set, 
obtained using the sets of data indicated by colour coding~\cite{EMMSY}.}
\end{figure}

Fig.~\ref{fig:interplay1} illustrates an example of the interplay between the
different constraints on a subset of operator coefficients, namely $\{ C_{tH}, C_{tG}, C_G, C_{HG} \}$. 
The Higgs data provide only weak constraints (yellow shading) before the $t \bar t H$ 
measurements are included (green shading), and substantial further improvements are provided
by including all the top data (pink shading) and four-fermion constraints (blue shading).
Two points to note are the strong correlations between pairs of operators, and the fact that
the Standard Model point is slightly outside the 95\% confidence level region. 
As discussed in~\cite{EMMSY}, this effect is largely driven by the measurement of the
$t \bar t$ cross section close to threshold, which favours a non-zero value of $C_G$.
However, multijet data impose a strong upper limit on $C_G$ when analyzed at quadratic
order (not considered here), suggesting that an alternative explanation of the apparent $t \bar t$
threshold discrepancy may be needed.

\begin{figure}[htb]! 
\hspace{-10mm}
\centering
\hspace{-10mm}
\includegraphics[angle=0,width=0.9\textwidth]{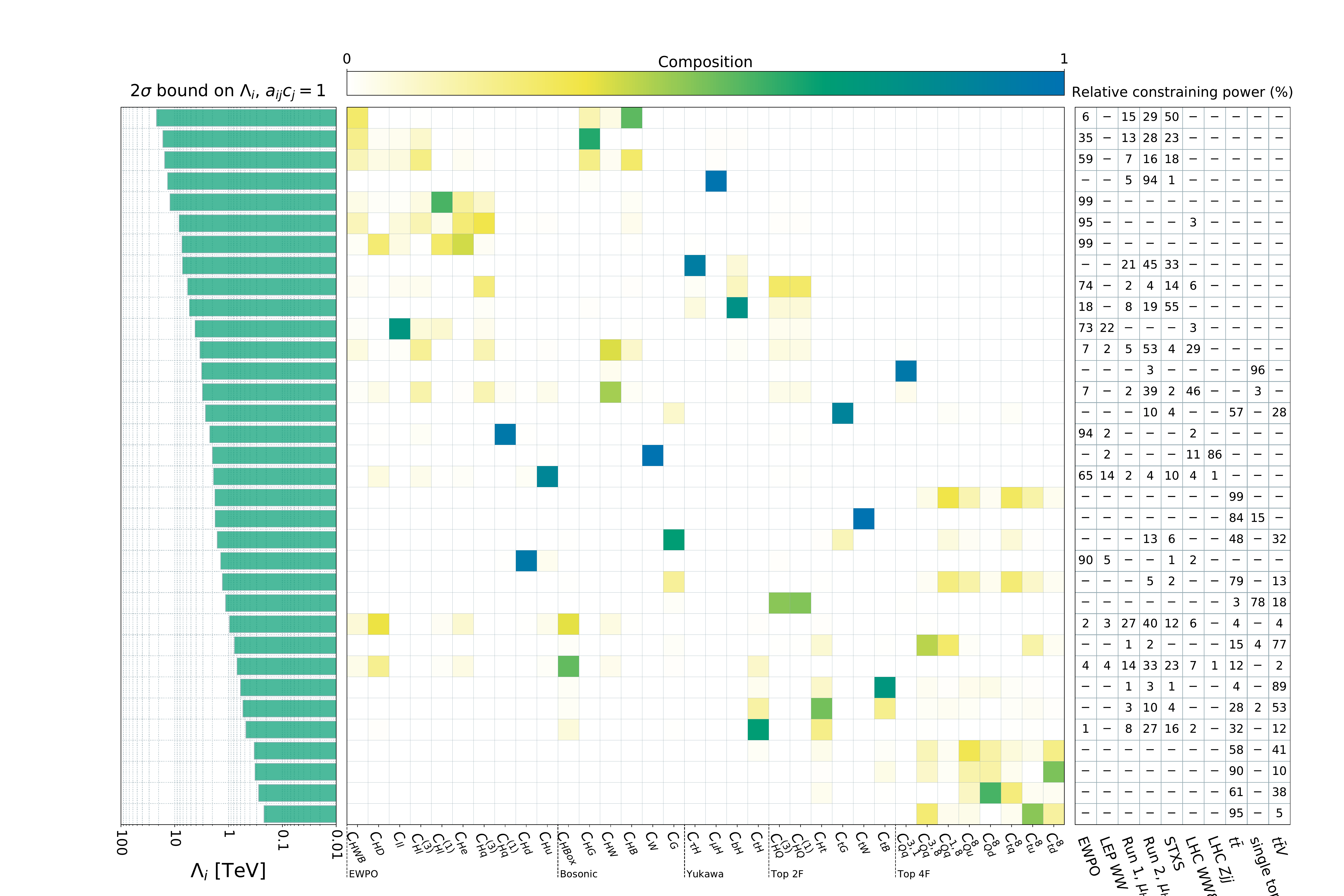}
\caption{\it \label{fig:eigenvectors} Constraints on the eigenvectors 
of the global likelihood function in a principal component analysis~\cite{EMMSY}.
The left panel shows the strengths of the global constraints on the
eigenvectors, the compositions of the eigenvectors
are colour-coded in the central panel, and the right panel shows the 
relative constraining powers of different data sets for each 
eigenvector. Cases with no significant constraining
power are indicated by {\rm ``-''}.}
\end{figure}

Fig.~\ref{fig:eigenvectors} shows the results of a principal component analysis of the
correlated constraints. The constraints on the scales $\Lambda$ of eigenvectors of the 
global likelihood function, normalizing the associated coefficient to unity,
are shown in the left panel, in order of decreasing sensitivity. The relative constraining powers of
different data sectors are shown in the right panel, and the compositions of the eigenvectors
are colour-coded in the central panel. We see that the least-constrained eigenvectors are only
constrained by data on $t \bar t$ production, single top production, and $t \bar t V$ production.
This reflects the fact that, as we already saw in Fig. ~\ref{fig:all_fit}, 
the precision of top data currently lags behind that in the
other data sectors.

\section{Single-Field Extensions of the Standard Model}

\begin{figure}[t!] 
\centering
\includegraphics[width=0.7\textwidth]{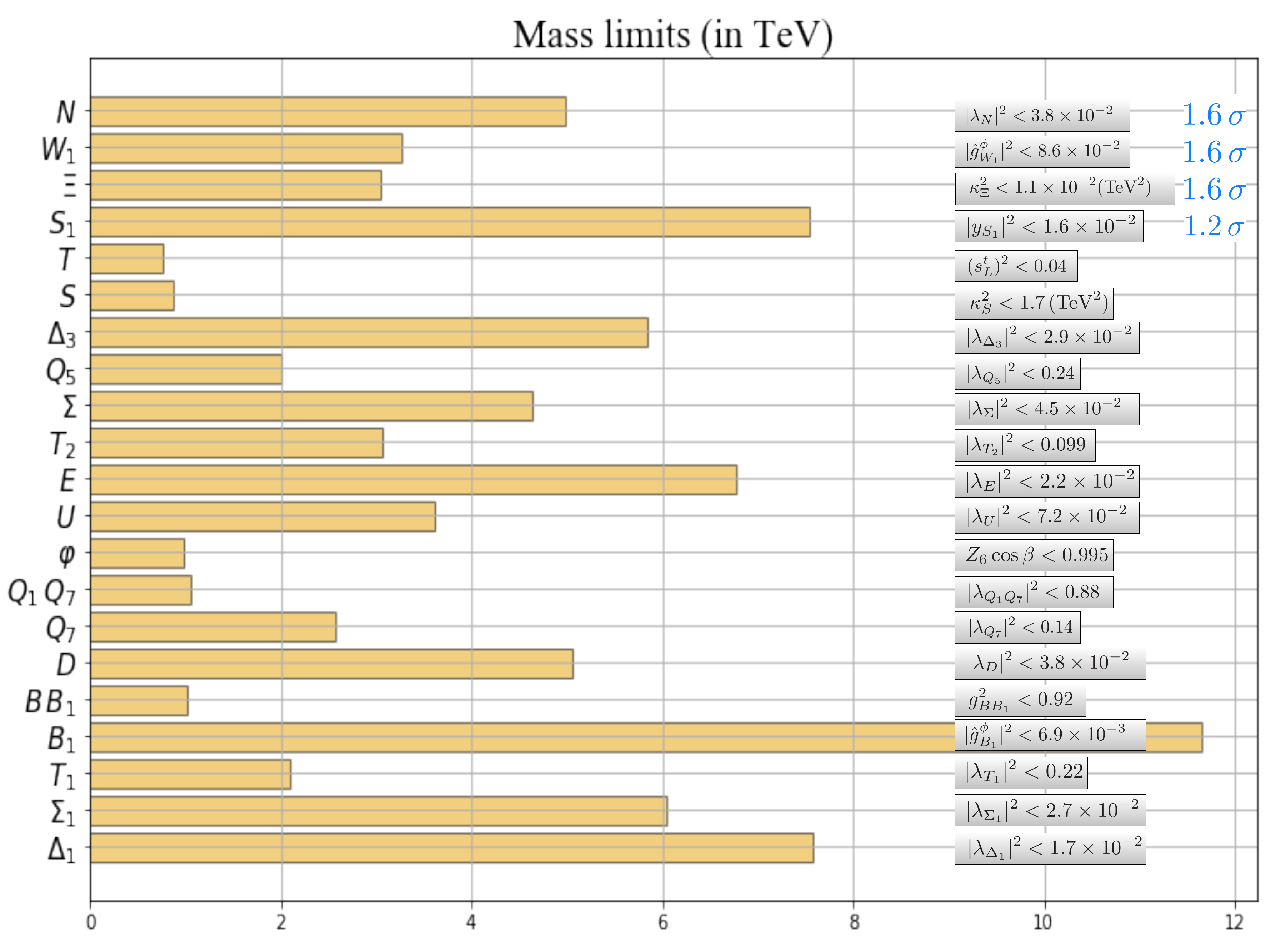}
\caption{\it Mass limits (in TeV) at the 95\% CL for single-field models assuming that
the corresponding couplings are unity, and limits on their couplings assuming masses of 1 TeV~\cite{EMMSY}.}
\label{fig:1Dlimits}
\end{figure}

As a first example of the constraints that our analysis places on specific scenarios for physics
beyond the Standard Model, we considered in~\cite{EMMSY} the single-field extensions of 
the Standard Model catalogued in the Table below. As indicated, each of these models may
be characterized by the mass of the additional field and a coupling constant. The tree-level contributions
of each of these single-field extensions to dimension-6 operator coefficients is given
in~\cite{dictionary}. These results can be combined with our global analysis to constrain the
mass of the additional field (assuming that the corresponding coupling is unity) or the coupling
(assuming that the corresponding mass is 1~TeV), as shown in Fig.~\ref{fig:1Dlimits}. As noted,
there is just a handful of instances where there is a deviation from the Standard Model that
exceeds 1 $\sigma$, none of which is significant. We find, in particular, that in all the possible
extensions with a single spin-zero field the particle mass must exceed about 800~GeV, while the
mass of any extra vector field must exceed about 1.5~TeV, if the corresponding coupling is unity.

  \begin{table}[h!]
\begin{center}
\resizebox{\columnwidth}{!}{%
\begin{tabular}{|c |c |c |c |c |c || c |c |c |c |c | c|} 
 \hline
Name & Spin & SU(3) & SU(2) & U(1) & Param. & Name & Spin & SU(3) & SU(2) & U(1) & Param. \\ [0.5ex] 
 \hline\hline
 $S$ & 0 & 1 & 1 & 0 & ($M_S$, $ \kappa_{S}$) &  $\Delta_{1}$ & $\frac{1}{2}$ & 1 & 2 & $- \frac{1}{2}$  & ($M_{\Delta_1}$,$\lambda_{\Delta_{1}}$)  \\ 
 \hline
 $S_{1}$ & 0 & 1 & 1 & 1 & ($M_{S_1}$,$y_{S_{1}}$) & $\Delta_{3}$ & $\frac{1}{2}$ & 1 & 2 & $- \frac{1}{2}$ & ($M_{\Delta_3}$,$\lambda_{\Delta_{3}}$)\\ 
 \hline
  $\varphi$ & 0 & 1 & 2 & $\frac{1}{2}$ & ($M_\varphi$,$ Z_{6} \cos \beta$) & $\Sigma$ & $\frac{1}{2}$ & 1 & 3 & 0  & ($M_\Sigma$,$\lambda_{\Sigma}$)\\ 
 \hline
  $\Xi$ & 0 & 1 & 3 & 0 & ($M_\Xi$,$\kappa_{\Xi}$)& $\Sigma_{1}$ & $\frac{1}{2}$ & 1 & 3 & -1 & ($M_{\Sigma_1}$,$\lambda_{\Sigma_{1}}$)\\ 
 \hline
   $\Xi_{1}$ & 0 & 1 & 3 & 1 & ($M_{\Xi_1}$,$\kappa_{\Xi_1}$) & $U$ & $\frac{1}{2}$ & 3 & 1 & $ \frac{2}{3}$ & ($M_{U}$,$\lambda_{U}$) \\ 
 \hline
$B$ & 1 & 1 & 1 & 0 & ($M_B$,$\hat  g_{H}^{B}$) & $D$ & $\frac{1}{2}$ & 3 & 1 & $- \frac{1}{3}$ & ($M_{D}$,$\lambda_D$) \\ 
 \hline
$B_{1}$ & 1 & 1 & 1 & 1 & ($M_{B_1}$,$g_{B_1}$)&$Q_{1}$ & $\frac{1}{2}$ & 3 & 2 & $ \frac{1}{6}$  &  ($M_{Q_1}$,$\lambda_{Q_1}$)\\ 
 \hline
  $W$ & 1 & 1 & 3 & 0 & ($M_W$,$ \hat{g}_{H}^{W}$)& $Q_{5}$ & $\frac{1}{2}$ & 3 & 2 & $- \frac{5}{6}$ & ($M_{Q_5}$,$\lambda_{Q_5}$)\\ 
 \hline
  $W_{1}$ & 1 & 1 & 3 & 1 & ($M_{W_1}$,$\hat{g}_{W_{1}}^{\phi}$)& $Q_{7}$ & $\frac{1}{2}$ & 3 & 2 & $\frac{7}{6}$ & ($M_{Q_7}$,$\lambda_{Q_7}$)\\ 
  \hline
   $N$ & $\frac{1}{2}$ & 1 & 1 & 0 & ($M_N$,$\lambda_{N}$)& $T_{1}$ & $\frac{1}{2}$ & 3 & 3 & $- \frac{1}{3}$ &  ($M_{T_1}$,$\lambda_{T_{1}}$)\\ 
  \hline
    $E$ & $\frac{1}{2}$ & 1 & 1 & -1 & ($M_E$,$\lambda_{E}$)& $T_{2}$ & $\frac{1}{2}$ & 3 & 3 & $\frac{2}{3}$  & ($M_{T_2}$,$\lambda_{T_2}$)\\ 
  \hline
  $T$ & $\frac{1}{2}$ & 3 & 1 & $\frac{2}{3}$ & ($M_T$,$s_{L}^{t}$)& $TB$ &
  $\frac{1}{2}$ & 3 & 2 & $\frac{1}{6}$ & ($M_{TB}$,$s^{t,b}_{L}$)\\ [1ex]
\hline
\end{tabular}
}
{\it Single-field extensions of the SM analyzed in~\cite{EMMSY}. \label{table:fields}}
\end{center}
\end{table}

\section{SMEFT Constraints on Light Stops}

Another example of a search for new physics beyond the Standard Model using the SMEFT is provided by
light stop squarks in supersymmetric models conserving $R$-parity~\cite{DEQY}. In this case there is no tree-level
contribution to any operator coefficients, but four coefficients receive one-loop contributions from
stop squarks~\cite{4stops}:
\begin{equation}
\label{stoperators}
\begin{split}
C_{HG} &= \frac{g_{s}^{2}}{12} \frac{h_{t}^{2}}{(4 \pi)^{2}}  \Big[ (1 + \frac{1}{12} \frac{c_{2 \beta} g^{'2}}{ h_{t}^{2}}) - \frac{1}{2} \frac{X_{t}^{2}}{m_{\tilde{t}}^{2}} \Big] \, ,\\
C_{HB} &=  \frac{17 g^{'2}}{144} \frac{h_{t}^{2}}{(4 \pi)^{2}}  \Big[ (1 + \frac{31}{102} \frac{c_{2 \beta} g^{' 2}}{ h_{t}^{2}}) - \frac{38}{85} \frac{X_{t}^{2}}{m_{\tilde{t}}^{2}} \Big] \, ,\\
C_{HW} &=  \frac{g^{2} }{16} \frac{h_{t}^{2}}{(4 \pi)^{2}}\Big[ (1 - \frac{1}{6} \frac{c_{2 \beta} g^{' 2}}{ h_{t}^{2}}) - \frac{2}{5} \frac{X_{t}^{2}}{m_{\tilde{t}}^{2}} \Big] \, ,\\
C_{HWB} &= - \frac{g g'}{24} \frac{h_{t}^{2}}{(4 \pi)^{2}}  \Big[ (1 + \frac{1}{2} \frac{c_{2 \beta} g^{2}}{ h_{t}^{2}}) - \frac{4}{5} \frac{X_{t}^{2}}{m_{\tilde{t}}^{2}} \Big] \, ,
\end{split}
\end{equation}
where $h_t$ is the top quark Yukawa coupling, the supersymmetric partners of the left- and right-handed
quarks are assumed to have the same mass $m_{\tilde t}$, $X_t$ is the off-diagonal term in their mass matrix, and
$c_{2 \beta} \equiv \cos 2 \beta$, where $\tan\beta$ is the ratio of supersymmetric Higgs vacuum expectation values.

\begin{figure}[t!] 
\centering
\includegraphics[width=0.5\textwidth]{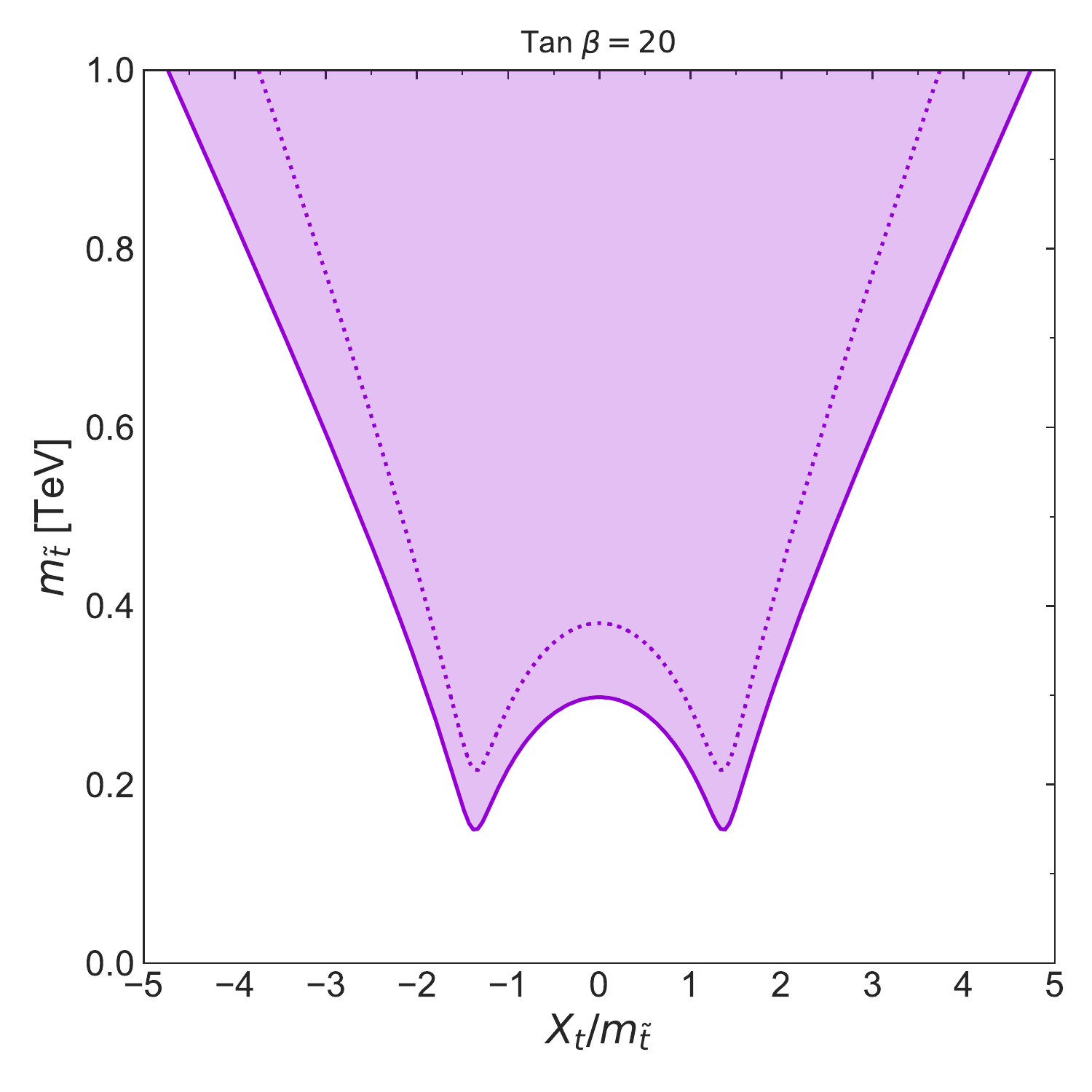}
\caption{\it SMEFT constraints on the stop parameters ($\frac{X_{t}}{m_{\tilde{t}}}, m_{\tilde{t}}$) for $\tan \beta= 20$~\cite{EMMSY}.}
\label{fig:MSSM}
\end{figure}

Fig.~\ref{fig:MSSM} shows the constraint in the $(X_t, m_{\tilde t})$ plane for $\tan \beta = 20$ that is
imposed by our SMEFT analysis. We see that values of $m_{\tilde t} \gtrsim 300$~GeV are favoured for most 
values of $X_t$. It is interesting to compare this indirect constraint with constraints from direct searches
for stop squarks at the LHC. These are quite complex, since the production and decays of stop squarks depend
on several othe supersymmetric model parameters, and some decay modes can be difficult to distinguish from
Standard Model backgrounds. On the other hand, the indirect search using the SMEFT is relatively
model-independent and of comparable strength.

\section{Model-Independent BSM Survey}

\begin{figure}[t!] 
\centering
\includegraphics[width=0.7\textwidth]{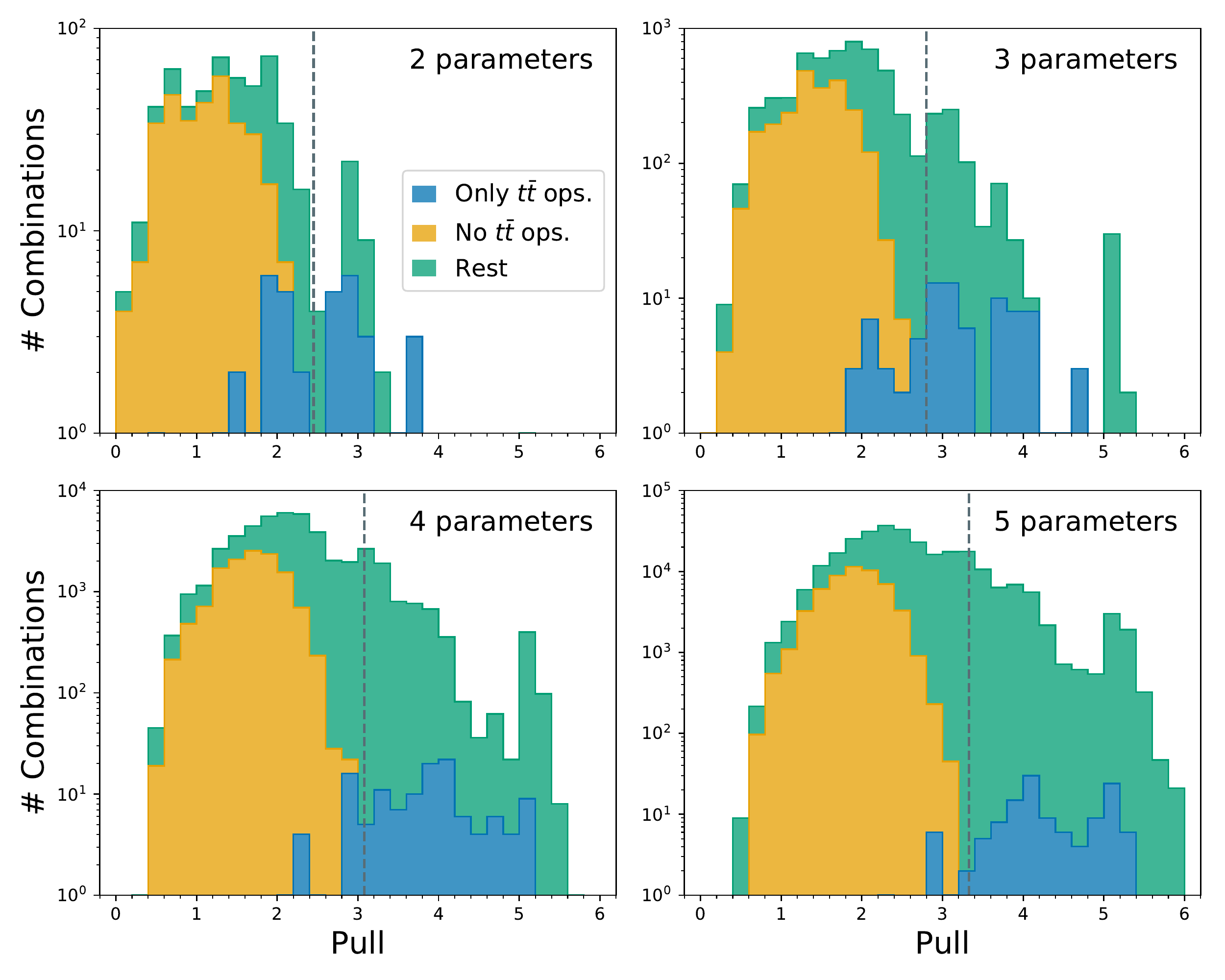}
\caption{\label{fig:pull_dist}
\it Stacked histograms of the pull distributions obtained in fits including 2 (upper left), 3 (upper right), 
4 (lower left) and 5 (lower right) dimension-6 SMEFT operators, divided into three categories: those including 
only operators affecting $t\bar{t}$ production (blue), those that do not affect $t\bar{t}$ production (orange), 
and the rest (green)~\cite{EMMSY}. The dashed vertical lines mark the expected 95\% ranges for the pull distributions.
}
\end{figure}

In addition to these searches for specific extensions of the Standard Model, we have also made a broad-band
search for generic models that contribute to arbitrary combinations of 2, 3, 4 or 5 operator coefficients.
For all of these 61, 5984, 46736 and 278256 combinations, we were able to calculate the corresponding
improvement in the global fit they provide in the forms of their pulls. The resulting pull distributions
are shown in Fig.~\ref{fig:pull_dist} as histograms for combinations containing only $t \bar t$ operators 
(blue), no $t \bar t$ operators (yellow) and both types of operator (green). The expected 95\% ranges for 
the pull distributions are shown as dashed vertical lines. As expected, some pulls lie outside these
range, particularly those containing $t \bar t$ operators. However, these tails of the distributions are
not excessively large, and we again see no significant evidence for new physics beyond the Standard Model. 

\section{Constraining Dimension-8 Terms in the SMEFT}

What about the contributions of SMEFT operators with dimension 8? These are difficult to isolate, as their
contributions are usually smothered by those of dimension 6. However, there are three interesting processes
where the leading SMEFT operators have dimension 8, providing novel windows on possible new physics beyond
the Standard Model.

\begin{figure}[h!]
\centering
\includegraphics[scale=0.35]{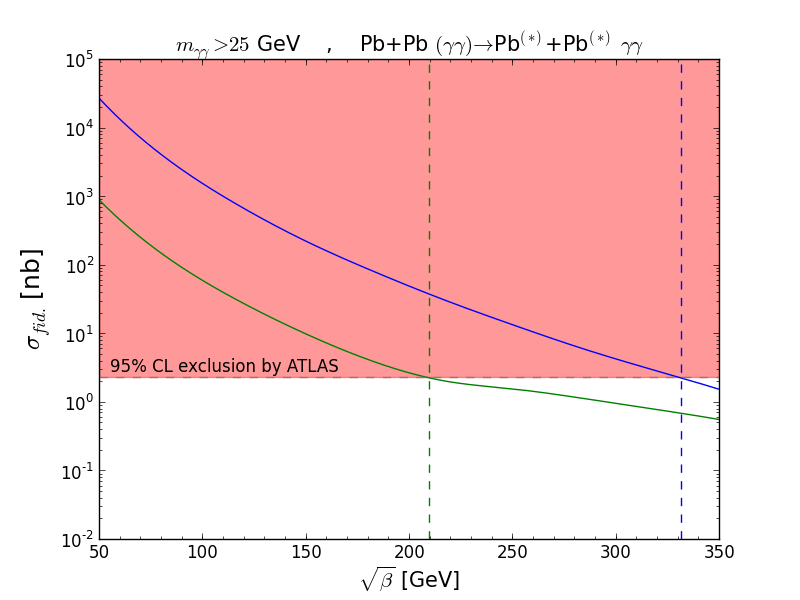}
\caption{\it The fiducial cross section for light-by-light scattering in relativistic heavy-ion collisions, $\sigma({\rm Pb} + {\rm Pb} (\gamma \gamma) \to
{\rm Pb}^{(*)} + {\rm Pb}^{(*)} \gamma \gamma)$ as a function of $M = \sqrt{\beta}$ in Born-Infeld theory obtained
with different theoretical assumptions (green and blue lines) is compared with the 95\% CL ATLAS upper limit~\protect\cite{ATLASgaga},
which excludes the range shaded pink~\cite{EMY}.}
\label{fig:betaconstraint}
\end{figure}

The first of these is light-by-light scattering, $\gamma \gamma \to \gamma \gamma$, which was first observed
in peripheral heavy-ion collisions at the LHC~\cite{ATLASgaga}. 
This process can arise from loop diagrams in the Standard Model, as was first
calculated by Heisenberg and Euler in the 1930s~\cite{HE}. Almost simultaneously, Born and Infeld~\cite{BI} suggested that there
might be nonlinearities in QED capable of generating light-by-light scattering, 
which they motivated by the idea that there should be a maximum
electromagnetic field, just as there is a maximum velocity, that of light. Some 50 years later it was shown that
such nonlinearities arise in string theory~\cite{Tseytlin}, and then it was shown in the 1990s
that in brane models they are due to brane
velocities being bounded by that of light~\cite{Bachas} - a remarkable vindication of the insight of Born and Infeld!
In the SMEFT language, the leading nonlinearity in the Born-Infeld Lagrangian is a term of dimension 8.
The LHC measurements were consistent with loop calculations in the Standard Model, and could be used to
constrain the mass parameter $M = \sqrt{\beta}$ appearing in Born-Infeld theory to be $\gtrsim 200$~GeV,
as seen in Fig.~\ref{fig:betaconstraint}~\cite{EMY}, which is the first significant experimental constraint on this parameter.

\begin{figure}[t!] 
\centering
\includegraphics[width=0.34\textwidth]{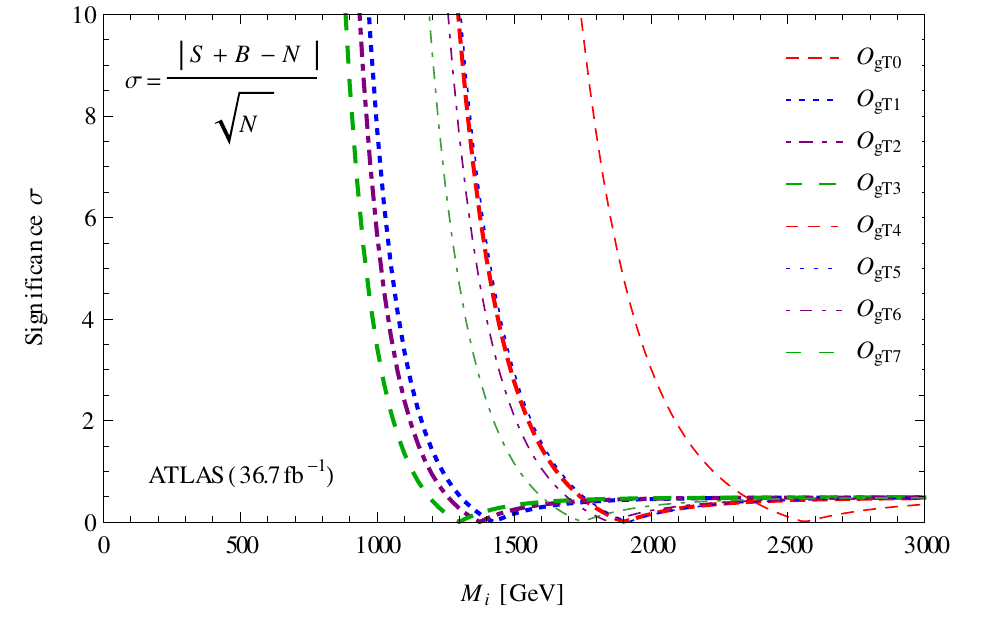}
\includegraphics[width=0.35\textwidth]{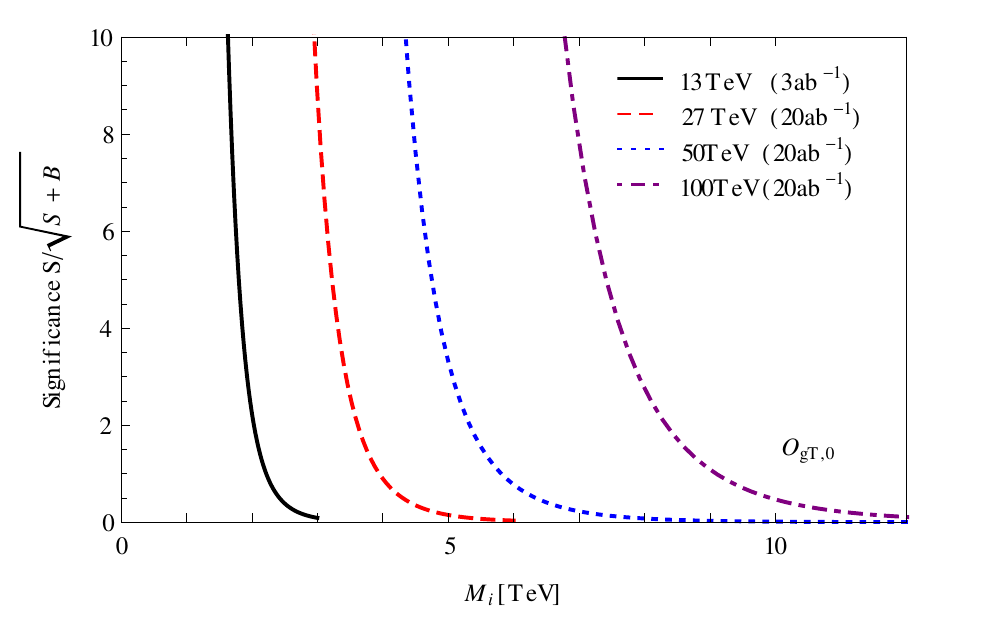}
\caption{\it Sensitivities to the new physics scales $\Lambda$ of the indicated dimension-8 
operators contributing to $gg \to \gamma \gamma$ in (a) early LHC Run~2 data and (b) possible
future data at the LHC and proposed $pp$ colliders~\cite{EGe}.}
\label{fig:gggaga}
\end{figure}

A second probe of dimension-8 terms in the SMEFT was provided by the process $gg \to \gamma \gamma$, which
could arise from Born-Infeld terms in the Standard Model~\cite{EGe}. The appearance of such terms
has been constrained by measurements of $\gamma \gamma$ final states in $pp$ collisions at the LHC~\cite{ATLASgaga2}. 
As seen in Fig.~\ref{fig:gggaga}, there are 8 dimension-8 operators that could contribute to this process.
The first analysis using early LHC Run~2 data shown in Fig.~\ref{fig:gggaga} provided lower limits
$\gtrsim 1000$~GeV on the scale parameters of all of these operators, and the sensitivity could be
improved to $\gtrsim 2$~TeV with data from HL-LHC and $\gtrsim 7$~TeV
at FCC-hh~\cite{EGe}.

\begin{figure*}[htb]!
\centering
\hspace*{-5mm}
\includegraphics[width=6cm,height=5cm]{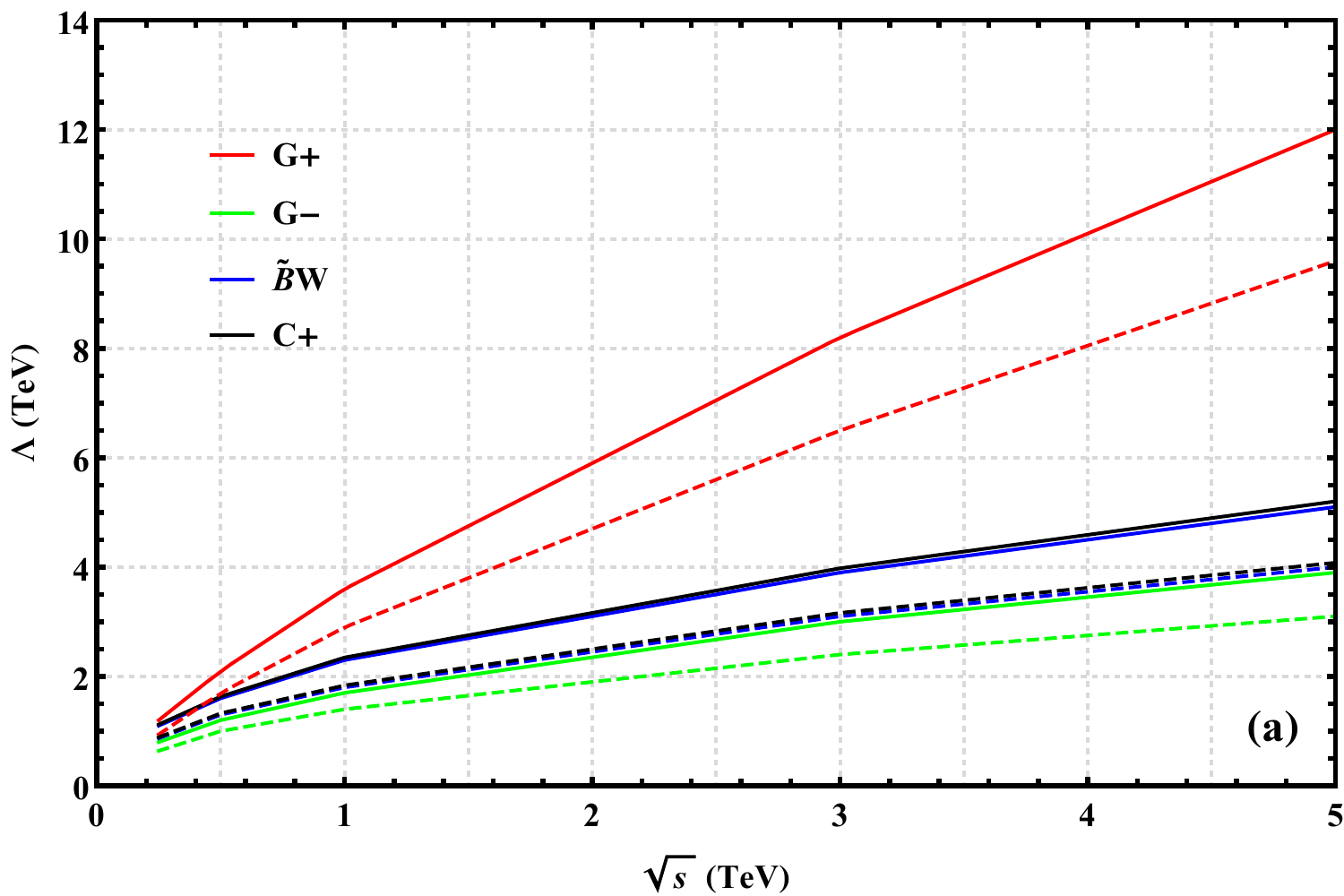}
\hspace*{2mm}
\includegraphics[width=6cm,height=5cm]{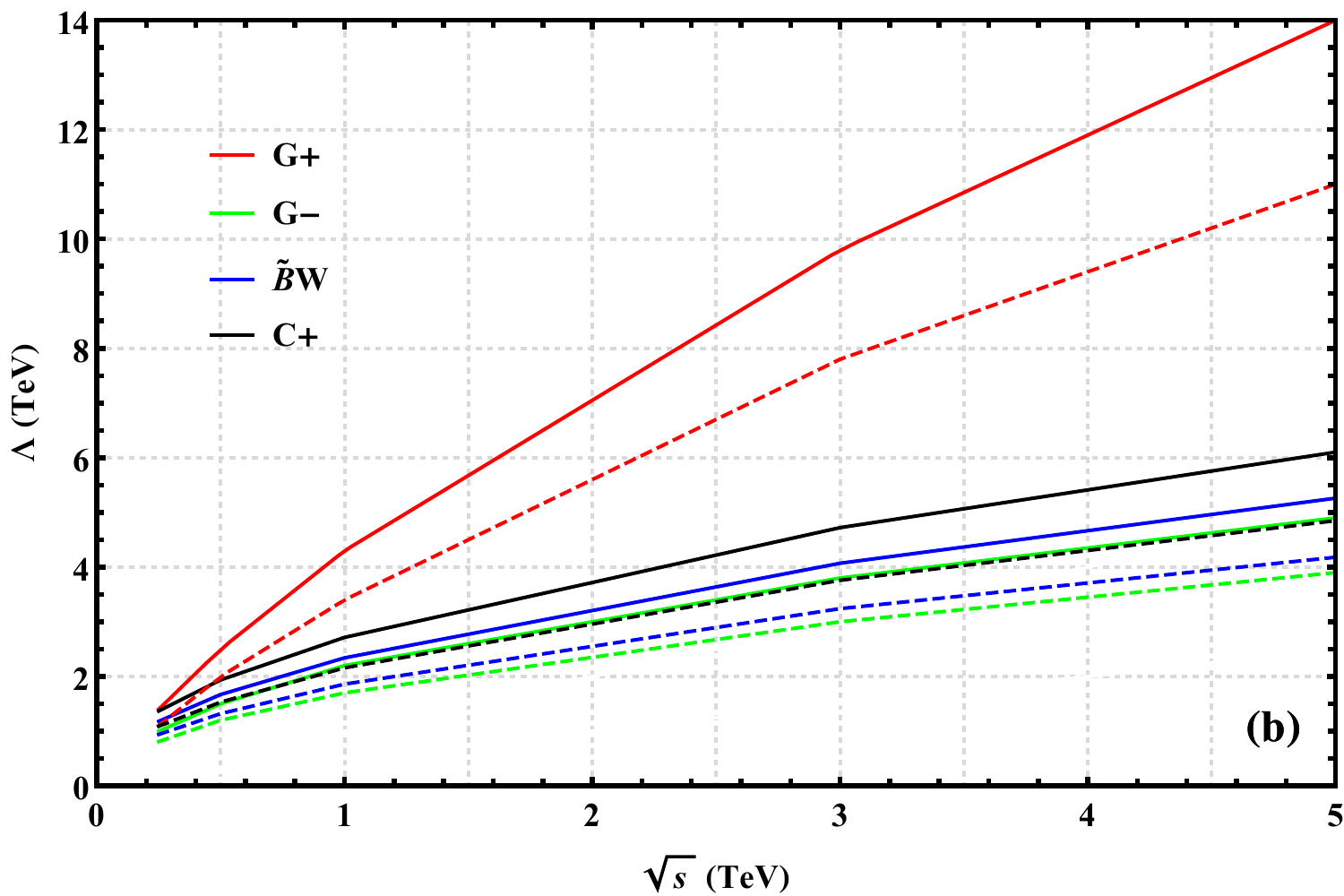}
%
\caption{\it Sensitivities to the new physics scales $\Lambda$ of the indicated dimension-8 
operators contributing to nTGVs, as functions
of the $e^+e^-$ collision energy $\sqrt{s}$ at the 2- and 5-$\sigma$ levels (solid and dashed lines),
assuming an integrated luminosity of 2~ab$^{-1}$ and (a) unpolarized and (b) polarized beams~\cite{EHX}.}
\label{fig:Zga}
\end{figure*}

Another opportunity to probe dimension-8 operators at future
accelerators is provided by measurements of neutral triple-gauge-boson
vertices (nTGVs), e.g., via the process $e^+ e^- \to Z \gamma$~\cite{EGHX, EHX}. The $ZZ\gamma$ and $Z \gamma \gamma$
nTGVs are absent in the Standard Model Lagrangian, and are not generated by dimension-6 SMEFT terms,
either. However, several types of nTGVs are generated by dimension-8 SMEFT operators, and generate
$Z \gamma$ final states in $e+e^-$ collisions with distinctive angular distributions that enable them
to be distinguished from Standard Model backgrounds. Fig.~\ref{fig:Zga} summarizes the potential
2- and 5-$\sigma$ sensitivities to the scales of dimension-8 operators provided by 2~ab$^{-1}$ of integrated
luminosity in unpolarized (left panel) and polarized (right panel)
$e^+ e^-$ collisions at a range of centre-of-mass energies from 250~GeV (corresponding
to FCC-ee, CEPC or the initial stage of the ILC) to 3 TeV (the target of CLIC) and above~\cite{EHX}. The possible
sensitivities reach into the multi-TeV range, depending on the operator considered, and improve if the $e^\pm$
beams are polarized.

\section{Summary}

Effective field theories have served particle physics well in the past, indicating
the direction to follow in constructing the electroweak sector of the Standard Model,
and also the structure of the gluon interactions underlying the strong force.
Many extensions of the Standard Model predict new particles and/or interactions at
scales beyond the masses of the known particles, which may show up first as
additional interactions in the Standard Model Effective Field Theory (SMEFT) that
indicate the direction in which new physics may lie. The most
prominent among these new interactions are likely to be those of dimension 6, but there
are also opportunities to search directly for possible new interactions of dimension 8.
This talk has summarized key features of a first global analysis of data from Run 2
of the LHC and previous experiments incorporating dimension-6 operators~\cite{EMMSY}, and how its
results can be used to constrain possible extensions of the Standard Model. In addition,
some of the opportunities for studying dimension-8 SMEFT operators have also been
mentioned.

\begin{figure}[t!] 
\centering
\includegraphics[width=0.5\textwidth]{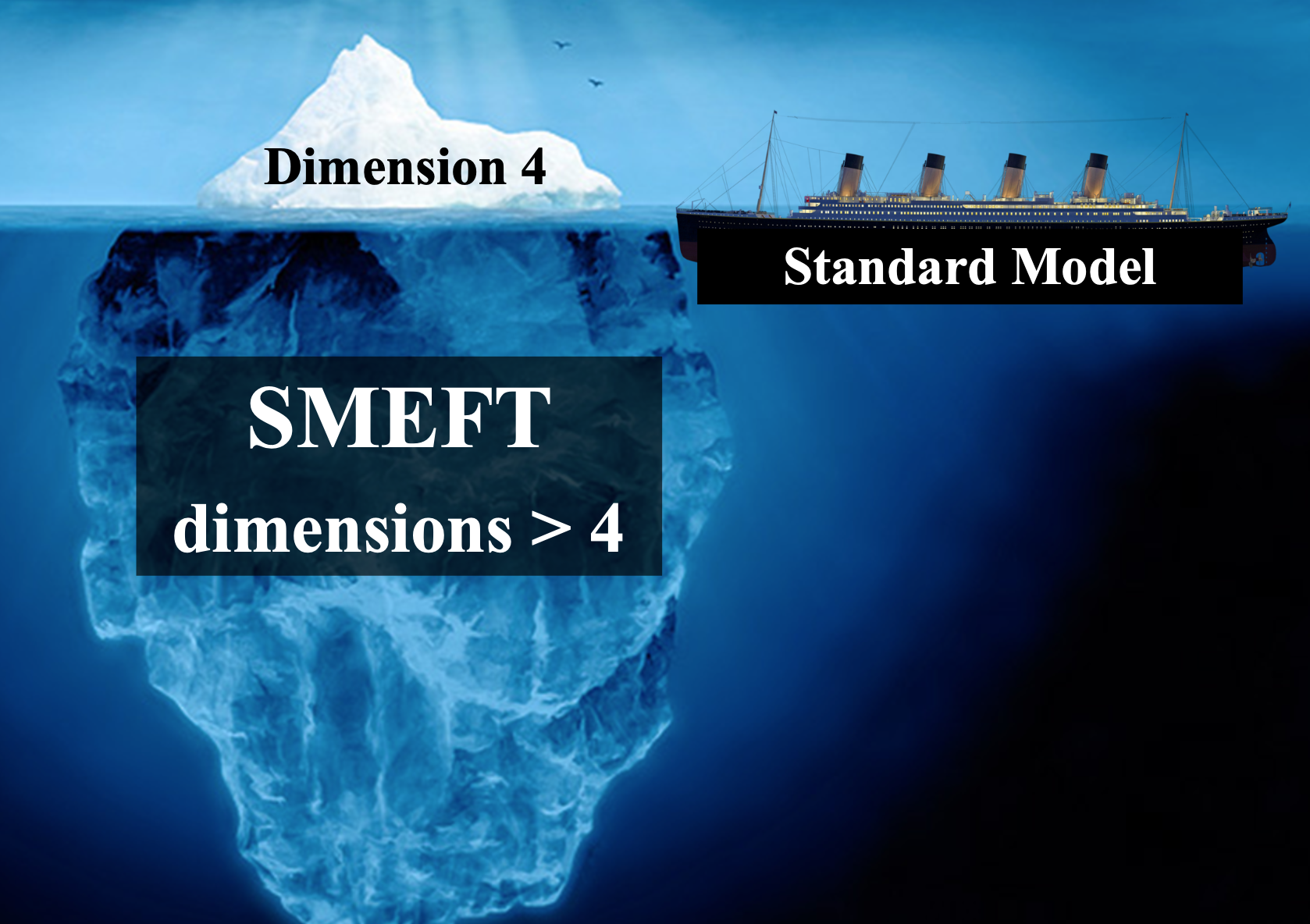}
\caption{\label{fig:Titanic}
\it Will the Standard Model soon be sunk by SMEFT interactions with dimensions $> 4$?
}
\end{figure}

So far, these SMEFT analyses have not provided any significant indication of possible
physics beyond the Standard Model. However, one may still hope that the far greater
integrated luminosity to be provided by future runs of the LHC will reveal significant 
deviations from the Standard Model, and indicate how Nature has chosen to go beyond it.
The prospects for the future are summarized in Fig.~\ref{fig:Titanic}. The dimension-4
Lagrangian of the Standard Model is only an infinitesimal part of the full set of
possible interactions between its particles that are represented by the SMEFT
interactions with dimensions $> 4$. So far, the Standard Model has been cruising along
untroubled, but it may soon collide with the SMEFT and sink, to be replaced by a more
complete theory.

\section*{Acknowledgements}
It is a pleasure to acknowledge my collaborators on the work described here: 
Ola Drozd, Shaofeng Ge, Hongjian He, Maeve Madigan, Nick Mavromatos,
Ken Mimasu, Chris Murphy, J{\' e}r{\' e}mie Quevillon, Veronica Sanz, Ruiqing Xiao and Tevong You.
The work described here was supported by the United Kingdom STFC via grant ST/T000759/1, 
and by the Estonian Research Council via a Mobilitas Pluss grant.

\bibliographystyle{unsrt}

\end{document}